\documentclass[twocolumn,prb,preprintnumbers,amsmath,amssymb]{aastex631} % 2 
\usepackage{amsmath} 
\usepackage{amsfonts} 
\usepackage{comment}
\usepackage{graphicx}
\usepackage{dcolumn}
\usepackage{bm}
\usepackage{rotating}

\begin{document}
\title{Early Solar System Turbulence Constrained by High Oxidation States \\of the Oldest Non-Carbonaceous Planetesimals}
\author{Teng Ee Yap}
\author{Konstantin Batygin}
\author{François L. H. Tissot}
\affil{\textit{The Isotoparium}, Division of Geological and Planetary Sciences, California Institute of Technology, Pasadena, CA 91125, USA.}

\begin{abstract}
\indent Early Solar System (SS) planetesimals constitute the parent bodies of most meteorites investigated today. Nucleosynthetic isotope anomalies of bulk meteorites have revealed a dichotomy between non-carbonaceous (NC) and carbonaceous (CC) groups. Planetesimals sampling NC and CC isotopic signatures are conventionally thought to originate from the “dry” inner disk, and volatile-rich outer disk, respectively, with their segregation enforced by a pressure bump close to the water-ice sublimation line, possibly tied to Jupiter's formation. This framework is challenged by emerging evidence that the oldest NC planetesimals (\textit{i.e.,} the iron meteorites parent bodies; IMPBs) were characterized by far higher oxidation states than previously imagined, suggesting abundant ($\gtrsim$ few wt.$\%$) liquid water in their interiors prior to core differentiation. In this paper, we employ a model for a degassing icy planetesimal (heated by $^{26}$Al decay) to map the conditions for liquid water production therein. Our work culminates in threshold characteristic sizes for pebbles composing the said planetesimal, under which water-ice melting occurs. Adopting a model for a disk evolving under both turbulence and magnetohydrodynamic disk winds, and assuming pebble growth is fragmentation-limited, we self-consistently translate the threshold pebble size to lower limits on early SS turbulence. We find that if NC IMPBs were “wet,” their constituent pebbles must have been smaller than a few centimeters, corresponding to typical values of the Shakura-Sunyaev $\alpha_{\nu}$ turbulence parameter in excess of $10^{-3}$. These findings argue against a quiescent SS disk (for $< 10$ AU), are concordant with astronomical constraints on protoplanetary disk turbulence, and suggest pebble accretion played a secondary role in building our rocky planets.
\end{abstract}

\section{Introduction}
\indent The formation of planetesimals via gravitational collapse of concentrated dust clouds in protoplanetary disks marks a critical transition point in planetary growth, straddling the gap between sub-meter pebbles and full-fledged planets thousands of kilometers in size (Youdin \& Goodman, 2005; Squire \& Hopkins, 2018; Klahr \& Schreiber, 2020). Most asteroids and meteorites investigated today constitute remnants of planetesimals from the early Solar System (SS). As such, they provide insight into the compositions and interior evolution of these primordial bodies, themselves reflective of provenance and prevailing conditions within the nascent circumsolar disk.\\ 
\indent The foremost of such conditions is the degree of turbulence, as it indirectly determines the characteristic Stokes number (a dimensionless measure of aerodynamic dust-gas coupling) and size of disk pebbles. The Stokes number in turn governs the efficiency of planetesimal formation and the dominant mode of subsequent planetary accretion (Birnstiel et al., 2012; Ormel, 2017; Batygin \& Morbidelli, 2022; Yap \& Batygin, 2024). At the macroscopic scale, turbulence manifests as an effective viscosity driving stellar accretion, and is conventionally quantified by the Shakura-Sunyaev $\alpha_{\nu}$ parameter (Shakura \& Sunyaev, 1973). In this work, we draw a connection between $\alpha_{\nu}$ and the oxidation states of the oldest planetesimals in the early SS. These planetesimals are the iron meteorite parent bodies (IMPBs), inferred to have accreted within 1 Myr after the condensation of the first solids in the SS [\textit{i.e.,} the Calcium-Aluminum-rich Inclusions (CAIs) prevalent in carbonaceous chondrites] (Kruijer et al., 2014, 2017; Hunt et al., 2018; Spitzer et al., 2021). \\
\begin{figure*} 
\centering
\scalebox{1.127}{\includegraphics{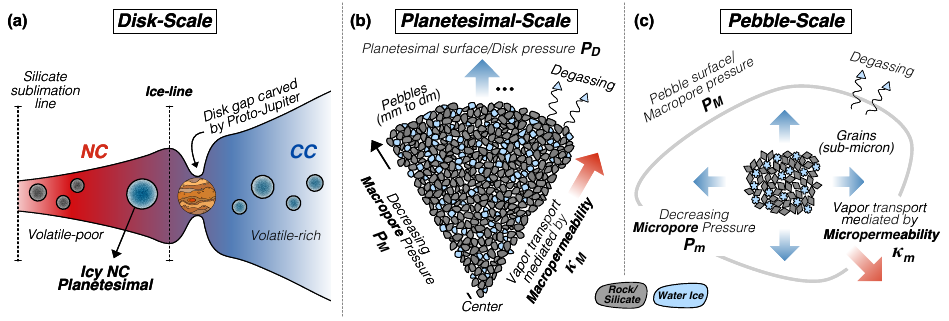}}
\caption{Illustrations of the icy NC planetesimal in consideration at the (a) disk-, (b) planetesimal-, and (c) pebble-scale. In (a), the planetesimal accreted beyond the ice-line now sits interior to it, such that its surface is permeable to water vapor loss. In (b), the packing of rocky and water-ice pebbles composing the planetesimal results in macropores, the pressure within which increases with depth. In (c), the packing of silicate and water-ice grains composing each rocky pebble results in micropores, the pressure within which increases with proximity to the pebble center (see \textbf{Section 2} \& \textbf{Table 1}).}
\label{fig:Figure 1}
\end{figure*}
\indent Over the past two decades, nucleosynthetic isotope anomalies in SS materials have revealed a dichotomy between non-carbonaceous (NC) and carbonaceous (CC) groups, testifying to the heterogeneous distribution of isotopically distinct presolar carriers in the SS disk (Warren, 2011; Budde et al., 2016; Burkhardt et al., 2019; Yap \& Tissot, 2023). Planetesimals sampling NC and CC isotopic signatures are thought to originate within and beyond the water-ice sublimation line (hereafter denoted the “ice-line”), respectively, with the early and prolonged ($\sim$ few Myr) separation of the two reservoirs typically attributed to a pressure bump just beyond the said line, possibly associated with Jupiter's formation (Kruijer et al., 2017; Brasser \& Mojzsis, 2020). The notion that NC planetesimals formed “dry” (\textit{i.e.,} absent of ice) has long been underpinned by the volatile-poor nature of NC meteorites and planetary bodies (\textit{i.e., }Earth, Mars, the Moon) relative to CC chondrites (\textit{e.g.,} Urey \& Craig, 1953; Warren, 2011; Marty 2012). This is reinforced by the higher Fe/Ni ratios observed in NC iron meteorites relative to their CC counterparts, thought (\textit{prima facie}) to reflect the scarcity of water in NC IMPBs, resulting in minimal sequestration of Fe as FeO in IMPB mantles following core segregation (Rubin, 2018; Spitzer et al., 2021; Hilton et al., 2022). \\
\indent While appealing for its simplicity, this framework is challenged by emerging evidence that NC IMPBs were characterized by oxidation states far higher than previously imagined, corresponding to $>$ few wt.\% liquid water therein. In particular, relying on Fe/(Ni, Co) ratios in NC and CC iron meteorites (assumed to be chondritic in their bulk parent bodies), Grewal et al. (2024) recently argued that NC IMPBs were only modestly reduced relative to their CC counterparts (oxygen fugacity greater than 3 log units below the Iron-Wüstite buffer). This paradigm shift was perhaps foreshadowed by the high oxidation state previously inferred for IVA (NC) iron meteorites, based on their Cr elemental and isotopic compositions (Bonnand \& Halliday, 2018). It is additionally supported, albeit indirectly, by the substantial presence of highly and moderately volatile elements (HVEs; \textit{e.g.,} C, N; Grewal et al., 2023; MVEs; \textit{e.g.,} S, Ge; Chabot et al., 2004; Goldstein et al., 2009) in several NC iron meteorite groups. Moreover, absorption features indicative of hydrated/oxidized silicates and molecular water have been observed for E- and S-type asteroids (\textit{e.g.,} Rivkin et al., 1995; Busarev, 2002; Arredondo et al., 2024), for which NC aubrites and ordinary chondrites (OC) are inferred analogs, respectively. \\
\indent An abundance of liquid water in the earliest NC planetesimals lends credence to models wherein both NC and CC planetesimals form at the forefront of an outward-migrating or periodically shifting ice-line, the latter reflecting variable stellar accretion (Voryobov, 2009; Lichtenberg et al., 2021). Alternatively, NC planetesimals formed beyond a stagnant ice-line over long timescales could have been scattered inwards by (proto-) Jupiter, with excited orbital eccentricities that are subsequently damped via gas drag (Adachi et al., 1976). Motivated by a \textit{supposed} need for “dry” NC planetesimals, Zhang (2023) recently developed a model for ice sublimation and vapor degassing therein. Notably, they showed that for icy planetesimals interior to the ice-line, and thus with surfaces permeable to water vapor, constituent pebble sizes in the few-cm- to m-scale can facilitate vapor loss while keeping pore pressures below the triple point pressure of  $\simeq$6 mbar, preventing melting and instantaneous (geologically-speaking) aqueous alteration/oxidation.\\
\indent Inspired by this newfound inclination for “wet” NC planetesimals, we employ such a model to investigate the conditions necessary to \textit{induce}, not prevent, the melting of water ice. Our work culminates in threshold characteristic pebble sizes below which liquid water would have been present in NC IMPBs. These sizes are translated to lower limits on $\alpha_{\nu}$ in the early SS, which are discussed in the context of empirical constraints on $\alpha_{\nu}$ from T-Tauri disks (\textit{e.g.,} ALMA observations; see \textbf{Section 5.1}) and how the SS planets came to be. 
\section{Model Overview: \\A Degassing Icy Planetesimal}
\indent Envision a sub-spherical planetesimal, newly-formed beyond the ice-line and comprising both rocky and water-ice pebbles. In the event that this icy planetesimal falls inward of the ice-line (\textbf{Fig. 1a}), its surface will be permeable to water vapor. As the decay of the short-lived radionuclide \textsuperscript{26}Al (\textit{e.g., }Macpherson et al., 1995; Hevey \& Sanders, 2006) proceeds, the internal temperature of the planetesimal will rise until it reaches the water triple point value of $\simeq$273 K. At this stage, \textit{depending} \textit{on pore gas pressures}, water-ice will either melt or sublimate. That is, energy from further decay will be dedicated to either the latent heat of melting or that of sublimation. We seek the region of parameter space in which the former is the case. Here, we provide an outline of the model adopted for the icy planetesimal, leaving an in-depth derivation of its constituent equations to \textbf{Appendix I}. This model builds on that from Zhang (2023). 

\subsection{Planetesimal Structure}
\indent In our model, the icy planetesimal of radius $R$ constitutes a homogeneous aggregate of rocky and water-ice pebbles with a characteristic size (radius) $a_p$. The packing of these pebbles results in a \textit{macroporosity} $\phi_M$. The rocky pebbles have a bulk density $\rho_r$ and themselves constitute a homogeneous mixture of silicate and water-ice \textit{grains} with characteristic size $a_g$. The packing of these grains results in a \textit{microporosity} $\phi_m$. Present-day astromaterials (\textit{e.g.,} chondrites) serving as analogs to the rocky pebbles have a ``anhydrous" microporosity $\phi_X$  ($X$ denoting the analog; \textit{e.g.,} OC), related to $\phi_m$ by the ice-to-silicate mass ratio $f$ at the time of planetesimal formation. In the case where rocky pebbles are devoid of ice (\textit{i.e.,} \textit{$f\simeq 0$}), $\phi_m\simeq \phi_X$. Note that $\rho_r$ is not equivalent to the bulk density of the assumed analog $\rho_X$, which does not account for the presence of ice. Instead, $\rho_r$ is given by $\rho_X (1+f)$.  The silicate grains have a density $\rho_s$, while water-ice pebbles and grains have a density $\rho_i$. An illustration of the icy planetesimal structure as described is given in \textbf{Figs. 1b \& 1c}.\\
\indent With the exception of $R$, constraints for the above parameters are obtained through experimental efforts. Hydrodynamic instabilities in protoplanetary disks concentrate pebbles in clouds that subsequently collapse by self-gravity to form planetesimals. Theoretical criteria for such collapse coupled with numerical simulations suggest planetesimal initial mass functions centered at \textit{R} on the order of $\sim$100 km (Klahr \& Schreiber, 2020; Gerbig \& Li, 2023), consistent with the sizes of Main Belt asteroids inferred to have evaded over four billion years of collisional evolution (between $\simeq35$ and $105$ km), constituting a population of “original” planetesimals (Delbo et al., 2017). Here, we allow $R$ to vary between 10 to 250 km, covering the range of radii inferred from IMPBs from iron meteorite/core cooling rates (Qin et al., 2008; see \textbf{Section 3.2}). The characteristic pebble size $a_p$ is treated as either a free parameter or model output. The characteristic grain size $a_g$, unless stated otherwise, takes on a fiducial value of $\simeq$100 nm, as determined for matrix grains in the C2 ungrouped chondrite Acfer 094 (Bland et al., 2009). Regarding $\phi_M$, we note that random close packing of equal-sized spherical particles results in a minimum porosity of $\simeq$0.36 (\textit{e.g.,} Scott \& Kilgour, 1969). The constituent pebbles of a planetesimal are, of course, neither perfectly spherical nor defined by a single size. Variations in pebble shape and/or size facilitate lower porosities, owing respectively to a greater number of rotational degrees of freedom in packing and the filling of interstitial spaces between larger pebbles that would otherwise be voids (\textit{e.g.,} Donev et al., 2004; Zhang et al., 2006). We let $\phi_M$ range between $\simeq$ 0.2 and 0.4. \\
\indent The bulk, rocky pebble density $\rho_r$, silicate grain density $\rho_s$, and present-day microporosity $\psi$ are dependent on the assumed analog material. In this work, we explore two endmember analogs: (i) OC and (ii) high-porosity (\textit{i.e.,} low thermal inertia) B-type asteroid Bennu surface rocks investigated \textit{in situ} by the OSIRIS-REx spacecraft that do not appear to be represented in the present meteoritic record, likely being too friable to survive atmospheric entry (Rozitis et al., 2020). For OC (mean of H, L, and LL falls), $\rho_{OC}$ and $\rho_s$ are $\simeq$3300 and 3600 kg/m\textsuperscript{3}, respectively, with $\phi_{OC}\simeq$0.07 (Consolmagno et al., 2008). For Bennu surface rocks (henceforth denoted “BA” for Bennu Analog), we take the average bulk and grain densities of CI and CM chondrites (lowest densities of all CC groups), yielding  $\rho_{CI/CM}$ and $\rho_s$ of  $\simeq$ 1900 and 2700 kg/m\textsuperscript{3}. respectively. We assume $\phi_{BA}\simeq$0.45 for BA. The water-ice pebble/grain density $\rho_i \simeq$ 1000 kg/m\textsuperscript{3}.  At the pebble-scale, $f$ is limited by $\rho_s$, $\rho_i$ and $\phi_X$ (see \textbf{Appendix I}). For simplicity, we set $f$ to half its maximal value (anti-correlated with $\phi_X$), such that $\phi_m$ for the OC analog and BA are $\simeq$ 0.04 and 0.23, respectively (\textit{i.e.,} $\phi_X$/2). Accordingly, $\rho_r$ [recall $\simeq \rho_X (1+f)$] for OC and BA are $\simeq$ 3368 and 2190 kg/m$^3$. \textbf{Table 1} summarizes the viable ranges and/or fiducial values for the structural parameters discussed. 

\begin{figure} 
\scalebox{0.93}{\includegraphics{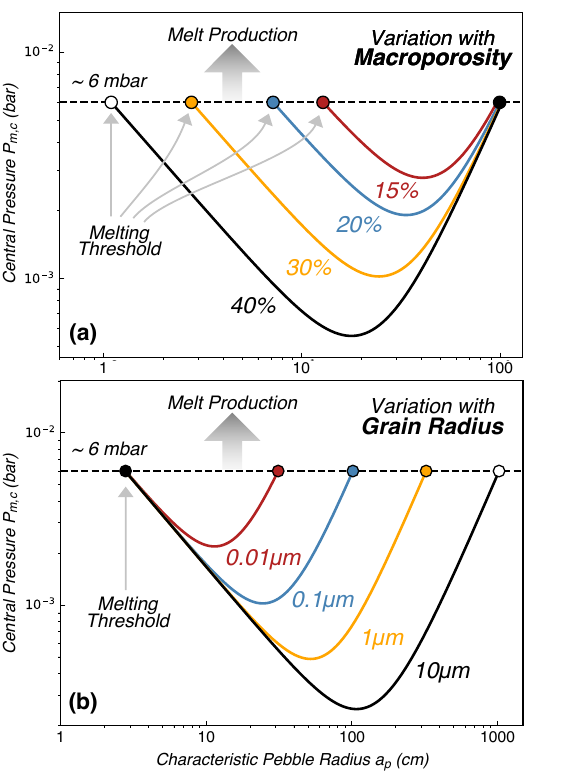}}
\caption{Plots of pressure at the centermost point in the planetesimal $P_{m,c}$ as a function of characteristic pebble radius $a_p$ for different (a) macroporosities $\phi_M$ and (b) grain radii $a_g$. Here, we assume $R\simeq100$km, $t_f\simeq0$ Myr, $P_D\simeq 10^{-4}$ bar, and OC as pebble analogs. For a given choice of $\phi_M$ and $a_g$, there are two threshold $a_p$ at which $P_{m,c}$ exceeds the water triple point pressure of $\simeq 6$mbar, corresponding to two regimes in the model. The relevant threshold in the context of planetesimal formation is the smaller of the two (see \textbf{Section 3.1} for discussion).}
\label{fig:Figure 3}
\end{figure}
\begin{figure*} 
\centering
\scalebox{0.8}{\includegraphics{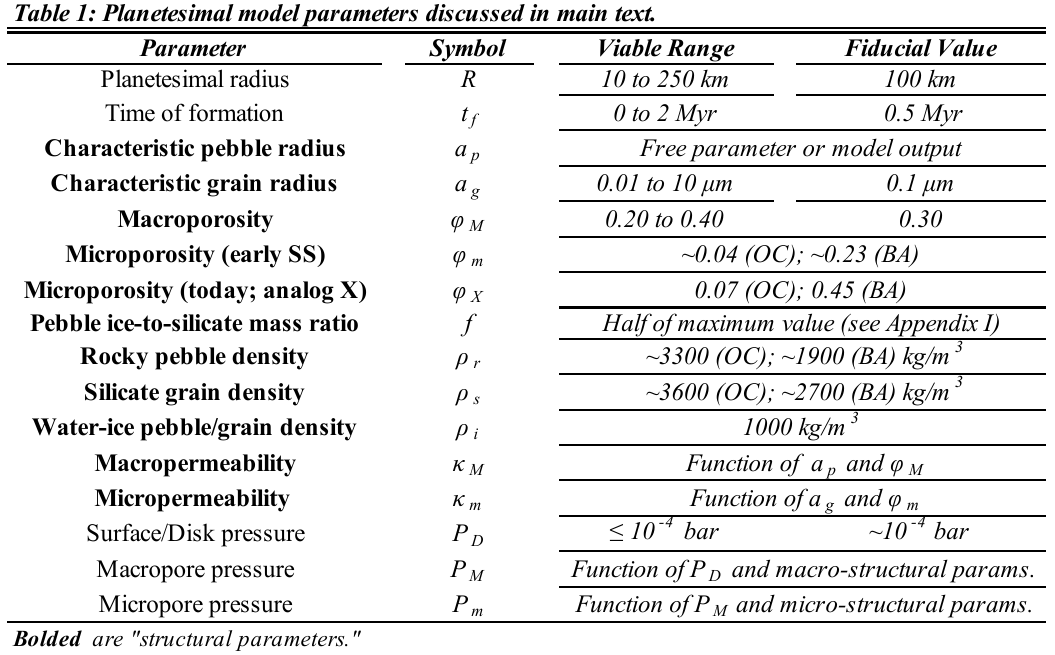}}
\end{figure*}
\subsection{Degassing Driven by Pore Pressures}
\indent The vapor production rate within the icy planetesimal (per unit mass) is a function of its formation time $t_f$ following the condensation of CAIs, serving as a proxy for \textsuperscript{26}Al activity. Vapor loss from the rocky pebbles constituting the planetesimal is driven by the pressure gradient between \textit{micro-} and \textit{macro-}pores (\textit{i.e.,} pore spaces \textit{within} and \textit{between} them). In turn, the pressure gradient between macropores and the planetesimal surface drives vapor loss to the protoplanetary disk (Zhang, 2023). The micropore pressure $P_m$ required to balance vapor production in a rocky pebble at $\simeq$273 K depends on the macropore pressure $P_M$, itself dependent (albeit insignificantly; see \textbf{Section 3}) on the disk/surface pressure $P_D$. The key parameter that relates $P_m$ to $P_M$ is the \textit{micro-}permeability $\kappa_m$, dependent on the pebble-scale structural parameters (\textit{i.e.,} $a_g$ \& $\phi_m$). Likewise, the \textit{macro-}permeability $\kappa_M$, dependent on the planetesimal-scale structural parameters (\textit{i.e.,} $a_p$ \& $\phi_M$), relate $P_M$ to $P_D$.  These permeabilities may be thought of as the effective cross-sectional area of the two “conduits” between micropore, macropore, and disk. \\
\indent The expression for $P_m$ at some radial distance $r_m$ from the center of a pebble, itself at a radial distance $r_M$ from the center of the planetesimal, takes the form of a quadratic (see \textbf{Appendix I}). The highest $P_m$ is realized at the heart of the rocky pebble sitting at the center of the planetesimal (\textit{i.e.,} $r_m\simeq r_M\simeq 0$). As such, the threshold for water-ice melting, however expressed, is determined with reference to $P_m$ at that point, denoted $P_{m,c}$. Having defined the two permeabilities, we can evaluate the $P_{m,c}$ required to balance vapor production. In particular, if  $P_{m,c}$ exceeds the water triple point pressure of $\simeq$ 6 mbar, melting occurs. 

\subsection{Underlying Model Assumptions}
\indent Before implementing our model just described, it is proper to consider caveats concerning its simplifying assumptions, which mark fertile ground for future developments. The first and foremost of these assumptions is that liquid water was the principal cause for high oxidation states in NC IMPBs. Occam's razor certainly suggests so, given that water has long been recognized as the primary oxidizing agent in early planetesimals (\textit{e.g.,} Ciesla \& Cuzzi, 2006; Grossman et al., 2012). Moreover, in the sole presence of water ice and/or vapor, alteration rates are expected to be negligibly small (Zhang, 2023). This is supported, for instance, by the preservation of abundant presolar silicate grains in volatile rich comets (\textit{e.g.,} 81P/Wild 2; Floss et al., 2013; Nguyen et al., 2023). In the event that oxidation is rapidly induced by water vapor, it would conceivably be confined to thin veneers that shield the bulk silicate material from further oxidation. The formation of such veneers has been investigated, and evidenced, in experiments of iron/alloy oxidation with water vapor, albeit at pressures and temperatures much higher than those relevant to our model (Fry et al., 2002; Bertrand et al., 2010; Yuan et al., 2013). While we proceed with our assumption, we stress that vapor-rock interactions should be explored in greater detail, and considered in future works. In addition to water (as either liquid or vapor), there exists alternative, and not mutually exclusive, explanations for oxidized material in the early SS. Notably, Ebel \& Grossman (2000) showed that equilibrium between gas and condensates under significant dust enrichment can endow the latter with oxidation states in excess of that inferred for IMPBs. More recently, Charnoz et al. (2024) suggested that in the case of either rapid ($<$ 1 year) or low pressure ($< 10^{-6}$ bar) kinetic condensation, appreciable production of oxidized species can occur.\\
\indent An important yet subtle assumption of our model is that the NC IMPB moves inward of the ice-line (rendering its surface permeable to water vapor) before its interior pressure and temperature reach the phase boundary between solid and gas, at which point sublimation takes place. An expansion of our present work could (or should) consider the possibility of delayed transport to within the ice-line, and the implications on planetesimal structure and interior evolution resulting therewith.\\
\indent Regarding structure, two central assumptions deserve mention: the negligence of (i) distributions in pebble size and shape, hinted above, and (ii) compaction accompanying sublimation (\textit{i.e.,} creation of void space). The former assumption is necessitated by uncertainty in the said distributions. This uncertainty is, nonetheless, captured in the range of (characteristic) pebble/grain sizes and macro-/microporosities explored, and in treating size and porosity as independent parameters (while they are related, note that two agglomerates with different monomodal particle sizes can have the same porosity on sufficiently large spatial scales, despite no variations in particle shape). Moving on to (ii), we note that restructuring can be confidently neglected so long as the planetesimal is rock/silicate-dominated. At the planetesimal scale, unless the planetesimal comprises abundant pebbles of $\sim$ pure water-ice, void space generated by sublimation is negligible. Indeed, Zhang (2023) (on whose work our planetesimal model is based) did not consider such pebbles in the first place. As detailed in the Appendix, the model simply assumes that most of the energy produced via $^{26}$Al decay is dedicated to ice sublimation. This ice can reside wholly in micropores. If water-ice pebbles were abundant, it can be safely assumed that the planetesimal will shrink over time (\textit{i.e.,} shortening the ``macro-conduit"; see Section 3), though it is unclear how/if the macroporosity would change. Compaction (\textit{i.e.,} decrease in macroporosity) over time would lead to a need for higher macropore pressures to drive vapor out. At the same time, the decrease in vapor production rate (\textit{i.e.,} $^{26}$Al activity) and planetesimal radius would decrease required macropore pressures. Future work exploring cases with abundant water-ice pebbles could, for instance, invoke a layered planetesimal structure, where macroporosity is a function of both depth and time.\\
\indent As with the planetesimal scale, at the pebble scale, restructuring is insignificant unless rocky pebbles comprise substantial water-ice grains. In modeling the formation and interior evolution of icy planetesimals, Lichtenberg et al. (2021) assumed an ice-rock ratio of unity. It is worthwhile to compare this value to the minimum ice-rock ratio (\textit{i.e.,} solely ice in micropores) for our model planetesimal. Assuming fiducial values for our icy planetesimal listed in Table 1, and rocky pebbles akin to BA, its minimum ice-rock ratio (\textit{i.e.,} solely ice in micropores) is calculated to be $\phi_{m}\rho_{i}/(1-\phi_X)\rho_{s}\simeq 0.15$. If all micropores were filled with ice (not half as assumed), this value comes out to be $\simeq 0.30$ (see $f_{max}$ in Appendix), a factor of a few less than unity. Of course, the ice-rock ratio can be increased by invoking a more microporous analog for the rocky pebbles, representing a precursor to observed astromaterials today. We note that the pebble-scale ice-rock ratio plays only a marginal role in our results, so long as pebble sizes do not change appreciably (and rapidly, since vapor production diminishes over time; see above) with sublimation. As discussed below in Section 3, in the regime relevant for planetesimal formation, the \textit{gradient in macropore (not micropore) pressure is key.} \\
\indent While the pursuit of developments aimed at relaxing the above assumptions are worthwhile, we do not foresee them altering our first-order conclusions dramatically (see Section 4). 

\section{Threshold Pebble Radius for Melting }
\indent Having identified a lithological analog for the pebbles constituting the planetesimal, we are left with three structural parameters to explore: (i) the characteristic pebble radius $a_p$, (ii) the macroporosity $\phi_M$, and (iii) the surface pressure $P_D$. Given a disk model, the latter is a proxy for heliocentric distance. In the disk region of interest (close to the ice-line; $>$1 AU), $P_D\sim10^{-4}$ bar in a standard viscous disk model, and we adopt this as its fiducial value. Accordingly, $P_D$ bears an insignificant influence on $P_{m,c}$ (the planetesimal surface is effectively a vacuum). Here, we investigate the relationship between $P_{m,c}$, $a_p$, and $\phi_M$, showing how a threshold (maximum) \textit{a}\textit{\textsubscript{p}} for melting of water-ice emerges as a meaningful value to pursue (\textbf{Section 3.1}). We proceed to find the threshold $a_p$ as a function of $t_f$ and $R$ (\textbf{Section 3.2}). 
\subsection{Interplay between Structural Parameters}
\indent While it is reasonable to assume a characteristic grain size $a_g$ on the order of 100 nm regardless of the pebble analog chosen (OC or BA), varying $a_g$ illuminates an interesting regime of model behavior. In \textbf{Fig. 2}, we plot $P_{m,c}$ as a function of $a_p$ for \textbf{(a)} a range of $\phi_M$ (15 to 40\%) and \textbf{(b)} a range of $a_g$ (0.01 to 10 $\mu$m). We assume pebbles of OC composition, $R\simeq$100 km, and $t_f\simeq0$. The leading order feature of both plots is the concave form of $P_{m,c}$, outlining a limited range of pebble radii $a_p$ within which no melting of water-ice occurs. The minimum in $P_{m,c}$, given $\phi_M$ in \textbf{Fig. 2a} or $a_g$ in \textbf{Fig. 2b}, constitutes a boundary between two regimes. To its left (\textit{i.e.,} towards smaller $a_p$), the macropore pressure $P_M$ (at $r_M\simeq 0$) dominates $P_{m,c}$ as the distance between micropores at the pebble center and the pebble surface is negligible. In other words, the “micro-conduit” with permeability $\kappa_m$ is short, such that $P_{m,c}\simeq P_M$. As $a_p$ decreases, the $P_M$ required to drive degassing at a rate comparable to that of vapor production increases (since the macropermeability $\kappa_M$ decreases). This proceeds until the triple point pressure is reached. To the right of the $P_{m,c}$ minimum (\textit{i.e.,} towards larger $a_p$), the pebble acts as a long “micro-conduit.” In this case, $P_{m,c}$ is largely set by the need to drive vapor out of the pebble to the macropores. As $a_p$ increases, the $P_{m,c}$ required to accomplish this increases until the triple point pressure is reached. \\
\indent Knowledge of these two regimes allows us to understand the trends observed in $P_{m,c}$ with changes in $\phi_M$ and $a_g$. For \textbf{Fig. 2a}, as $\phi_M$ enters into the calculation of $\kappa_M$(\textit{i.e.,} the permeability of the dominating “conduit” at small $a_p$), its variance primarily affects the point at which melting occurs to the left of the  $P_{m,c}$ minimum. As  $\phi_M$ increases, $a_p$ must decrease to yield the same $\kappa_M$. Accordingly, the point at which $P_M\simeq P_{m,c}\simeq 6$ mbar shifts to smaller $a_p$. Increasing $\phi_M$ by a factor of a few, from $\simeq$ 15 to 40\%, leads to a decrease in the threshold $a_p$ by an order of magnitude, from $\simeq$ 15 to 1 cm. For \textbf{Fig. 2b}, $a_g$ enters into the calculation of $\kappa_m$(\textit{i.e.,} the permeability of the dominating “conduit” at large  $a_p$), such that its variance affects the point at which melting occurs to the right of the  $P_{m,c}$ minimum. As $a_g$ increases, the point at which  $P_{m,c}\simeq 6$mbar $>P_M$ is obtained for a greater length of the “micro-conduit” $a_p$. In particular, $\gtrsim$ an order of magnitude increase in the threshold $a_p$ accompanies the increase in $a_g$ by three order of magnitude, from $\simeq$ 0.01 to 10 $\mu$m.\\
\indent In planetesimals, the relevant/meaningful threshold $a_p$ at which water-ice melting occurs is the smaller of the two, to the left of the  $P_{m,c}$ minimum and in the cm- to $\sim 1$ dm-scale. Indeed, both astronomical observations (\textit{e.g.,} Testi et la., 2003; Ricci et al., 2010; Birnstiel, 2023 \& references therein) and dust coagulation models (\textit{e.g.,} Birnstiel et al., 2012; Batygin \& Morbidelli, 2022; Yap \& Batygin, 2024) infer an abundance of mm- to cm-sized pebbles in protoplanetary disks. Moreover, bearing in mind the eminent “meter-size barrier” to pebble growth from radial drift (Weidenschilling \& Cuzzi, 1993; Blum \& Wurm, 2008), it is improbable that pebbles larger than several decimeters characterize the size distribution in any disk region. Moving forward, we denote the threshold (\textit{i.e.,} maximum) characteristic pebble radius for the presence of liquid water in the planetesimal interior as $a_{p,th}$. 
\begin{figure*} 
\centering
\scalebox{0.97}{\includegraphics{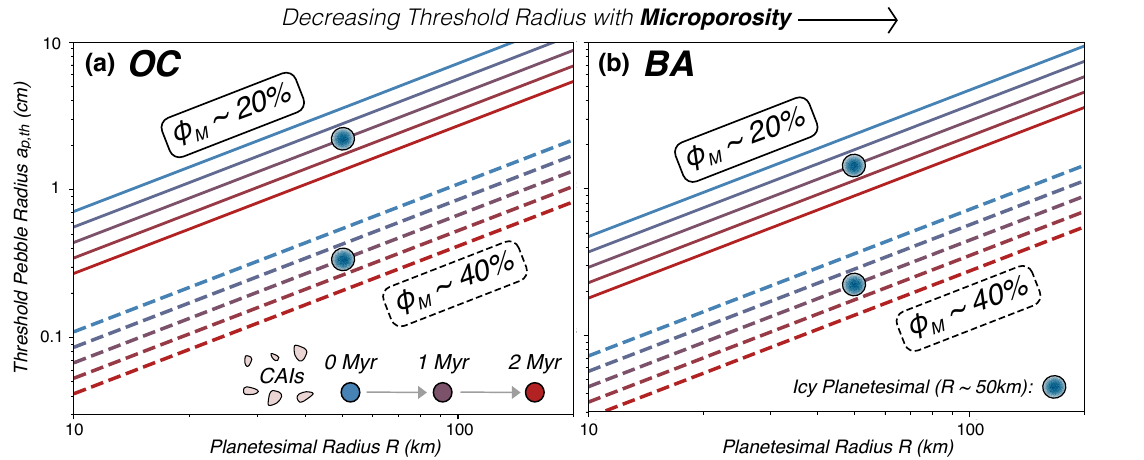}}
\caption{Plots of the threshold pebble radius $a_{p,th}$ for melting water-ice as a function of planetesimal radius $R$ assuming (a) OC and (b) BA as pebble analogs. As is clear, $a_{p,th}$ decreases with increasing formation time $t_f$, macroporosity $\phi_M$, and microporosity $\phi_m$, but increases with $R$. Points corresponding to $R\simeq 50$ km and $t_f\simeq 0.5$ Myr are designated with a blue marker (see \textbf{Section 3.2} for discussion).}
\label{fig:Figure 3}
\end{figure*}
\subsection{Probing Parameter Space}
\indent Having established the existence of a melting threshold $a_{p,th}$, we now account for its dependence on (i) planetesimal radius $R$, (ii) pebble microporosity $\phi_m$ (representing the analog chosen), and (iii) post-CAI formation time $t_f$, a proxy for the $^{26}$Al activity in the planetesimal's silicates. The activity directly relates to the vapor production rate, assuming all energy derived from decay is expended on ice sublimation (see \textbf{Appendix I}).  In \textbf{Fig. 3}, we plot $a_{p,th}$ as a function of $R$ assuming \textbf{(a)} OC and \textbf{(b)} BA pebbles, each for two $\phi_M$ (20 and 40\%) and $t_f$ spanning the $2$ Myr following CAI condensation. The trends observed with changes in $R$, $\phi_m$, and $t_f$ are easily understood given the intuition developed in the previous section. Starting with the former, as $R$ (\textit{i.e.,} the ``macro-conduit" length) increases, the cumulative pressure change in going from planetesimal surface to center increases, requiring larger pebbles to keep the planetesimal ``dry." An increase in $\phi_m$ corresponds to a decrease in the pressure gradient between macro- and micropores. This places a greater burden on the ``macro-conduit" to generate the difference between $P_{m,c}$ and $P_D$. This difference increases with decreasing $a_p$. Note that this effect is insignificant, as expected for the low-$a_{p}$ regime wherein $P_{m,c}\simeq P_M$ (see \textbf{Section 2.2}). Finally, the later a planetesimal with a given macropermeability $\kappa_M$ and radius $R$ forms, the smaller its vapor production rate, and the smaller the pressure gradient required to drive degassing. This means $a_p$ (and thus $\kappa_M$ for a set $\phi_M$) must decrease to induce melting, and so $a_{p,th}$ is lower. \\
\indent  At this stage, it behooves us to consider the values of $R$ and $t_f$ applicable to NC IMPBs. Metallographic cooling rates of NC iron meteorites (\textit{i.e.,} IVA, IIIE, IIAB, IIIAB, and IIC groups) suggests their parent bodies ranged in size from $R\simeq 10$ to $>200$ km (Qin et al., 2008; Goldstein et al., 2009). A fiducial $R\simeq 50$ to $100$ km is thus reasonable to assume. Regarding $t_f$, Hf-W core formation model ages of NC IMPBs range between $\simeq 1$ to $3$ Myr after CAIs (\textit{e.g.,} Spitzer et al., 2021; Hellmann et al., 2024). Using these ages as an anchor, thermal models assuming $^{26}$Al decay prompted planetesimal differentiation constrain the accretion time of NC (and CC) IMPBs to $<1$ Myr.  Assuming $R\simeq 50$ km and $t_f\simeq 0.5$ Myr (\textbf{Fig. 3}), we obtain $a_{p,th}$ between a few mm to cm across the range of $\phi_M$. For $R \gtrsim 100$ km, this range is shifted up by a factor of a few. 
\section{Constraints on Disk Turbulence}
\subsection{From planetesimal to disk model}
\indent Threshold (maximal) characteristic pebble radii $a_{p,th}$ for melting water-ice in the planetesimal interior can be translated to lower limits on disk turbulence, the typical range for which is $10^{-4}\lesssim\alpha_{\nu}\lesssim 10^{-2}$ (\textit{e.g.,} Armitage, 2020; Rosotti et al., 2023). This connection derives from the assumption that hit-and-stick pebble growth in disks is limited by collisional fragmentation, due to relative velocities arising from either turbulent motion or radial drift (Ormel \& Cuzzi, 2007; Birnstiel et al., 2012; Batygin \& Morbidelli, 2022; Yap \& Batygin, 2024). Relative drift velocities relate to turbulence as it facilitates angular momentum redistribution in the disk, driving the accretionary flow of gas and dust. The larger pebbles grow (\textit{i.e.,} the larger their Stokes numbers, $St$, grow), the greater their (turbulent and drift) velocity dispersion and collisional velocities. As such, depending on location in the disk and pebble composition (rock and ice differ in tensile strength, and thus fragmentation threshold velocities), a characteristic radius/$St$ is reached at which fragmentation and growth are at quasi-equilibrium. As planetesimals originate from pebble clouds, this characteristic radius is roughly equivalent to that in the icy planetesimal $a_p$. Indeed, steaming instabilities (or more generally, resonant drag instabilities) thought to form pebble clouds tend to do so with the largest pebbles in the disk locality (Youdin \& Goodman, 2005; Squire \& Hopkins, 2018). \textit{Greater $\alpha_{\nu}$ translates to greater collisional velocities for a given distribution of pebble $St$, in turn leading to smaller $a_p$ and ultimately higher pressures in the planetesimal.} An important assumption worth noting here is that disk pebbles do not grow substantially within clouds before their collapse (if they do, the threshold $\alpha_{\nu}$ obtained for the onset of melting remains a lower limit $\rightarrow$ pebbles need to be even smaller before partaking in gravito-hydrodynamic instabilities). \\
\indent The conversion of $a_{p,th}$ to a threshold (minimum) turbulence parameter $\alpha_{\nu,th}$ requires a disk model, with which we can position ourselves close to the ice-line and calculate $St$. Here, we provide a brief overview of such a model, and defer the reader to \textbf{Appendix II} for a more detailed explanation thereof (for an comprehensive description of the model, see Yap \& Batygin, 2024). The model adopted features a steady-state disk whose accretion is driven by turbulence and magnetohydrodynamic (MHD) disk winds (Tabone et al., 2022), and is built upon a well-established framework that self-consistently relates the disk surface density to its midplane temperature, macroscopic viscosity, etc. (Shakura \& Sunyaev, 1973; Lynden-Bell \& Pringle, 1974; Armitage, 2020). The region of interest is assumed to be sufficiently far within the characteristic outer radius of the disk such that stellar irradiation contributes negligibly to the midplane temperature (\textit{e.g.,} Chambers, 2009). Accordingly, we assume heating is derived from viscous shear alone. Barring constants that define the physical dimensions of the disk (\textit{e.g.,} the total disk mass), two free parameters specify the significance of turbulence in regulating pebble-to-planetary growth. The first of these is $\tilde{\alpha}$, quantifying the total torque on the disk and constituting a sum of $\alpha_{\nu}$ and an analogous parameter $\alpha_{DW}$ for MHD winds:
\begin{equation}
\tilde{\alpha} = \alpha_{\nu} + \alpha_{DW}.
\end{equation}
To specify the relative strength between turbulence and winds, we have $\psi$, defined by the ratio of $\alpha_{DW}$ to $\alpha_{\nu}$:
\begin{equation}
\psi = \alpha_{DW}/\alpha_{\nu}.
\end{equation}
As such, low and high $\psi$ correspond to turbulence and wind dominance, respectively. Having specified $\tilde{\alpha}$ and $\psi$, $\alpha_{\nu}$ is given by
\begin{equation}
\alpha_{\nu} = \tilde{\alpha}/(1+\psi).
\end{equation}
\indent With a disk model established, $a_{p,th}$ as obtained via the planetesimal model can be converted to a threshold $St$, denoted $St_{th}$ assuming the proper drag regime (at $\sim$ the ice-line; defined by $T\simeq 170K$). In the ensuing analysis, it was affirmed \textit{a posteriori} that the Epstein drag regime is applicable given the fiducial values adopted in both the planetesimal and disk models. This regime is valid so long as $a_{p,th}\lesssim 9\lambda_{mfp}/4$, where $\lambda_{mfp}$ is the mean free path of (predominantly $H_2$) gas molecules. The $St_{th}$ thus obtained cannot be converted directly to $\alpha_{\nu,th}$, as (i) there are two fragmentation barriers to pebble growth (see above), and it is unclear which is dominant \textit{a priori}, and (ii) the relationship between $\alpha_{\nu}$ and $St$ for fragmentation due to relative drift velocities is not analytically solved, involving as it does a quartic equation (see \textbf{Appendix II}). 
\begin{figure} 
\scalebox{1.25}{\includegraphics{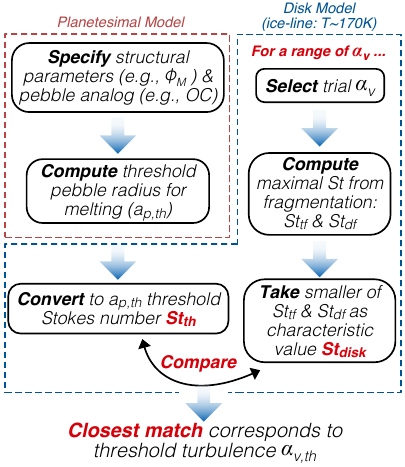}}
\caption{Flowchart displaying the procedure for computing the threshold turbulence parameter $\alpha_{\nu,th}$, involving an iterative process seeking to minimize the deviation between the threshold Stokes number $St_{th}$ (from $a_{p,th}$) and the characteristic Stokes number of disk pebbles limited by fragmentation, independent of the planetesimal model (see \textbf{Section 4.2} for details).}
\label{fig:Figure 4}
\end{figure}
\subsection{From disk model alone }
\indent \textit{Independent of the planetesimal model}, the characteristic $St$ of pebbles is defined at any given region in the disk through the competition between turbulent and relative drift fragmentation. These two barriers each set a maximal $St$ attainable by pebble undergoing collisional growth, the smaller of which constitutes the characteristic $St$, henceforth denoted $St_{disk}$. Put simply, pebble growth starting at the micron-scale is halted by the first fragmentation barrier it encounters. Underlying this framework is the assumption that growth operates at a sufficiently high efficiency such that most pebbles have $St\simeq St_{disk}$. The maximal $St$ values are denoted $St_{tf}$ and $St_{df}$, respectively, and so
\begin{equation}
St_{disk} = min[St_{tf},St_{df}].
\end{equation}
As with details of the disk model, the equations relating $\alpha_{\nu}$ to $St_{tf}$ and $St_{df}$ are provided in \textbf{Appendix II}. A key parameter in these equations is the fragmentation threshold velocity of icy grains $v_{f,ice}$, for which we adopt either $5$ or $10$ m/s, larger than that typically assumed for pure rock $v_{f,rock}\simeq 1$m/s (\textit{e.g.,} Armitage, 2020; Yap \& Batygin, 2024). Recent laboratory experiments suggests at low disk temperatures ($T< 150K$), icy pebbles have no advantage over their silicate counterparts interior to the ice-line in tensile strength (\textit{e.g.,} Gundlach et al., 2018; Musiolik \& Wurm, 2019). As such, it is conceivable that $v_{f,ice}$ peaks close to the ice-line before decaying to a value akin to $v_{f,rock}$ farther out (Yap \& Batygin, 2024). As the NC planetesimal under consideration is envisioned to form at or just beyond the ice-line, it is safe to assume $v_{f,ice} >$ 1 m/s. Owing to the difference in density between the pebble analogs OC and BA, they ought to be characterized by different $v_{f}$. Moreover, $v_f$ in general depends on pebble size and temperature (\textit{e.g.,} Benz \& Asphaug, 1999; Leinhardt \& Stewart, 2009; Gundlach et al., 2018). These caveats are omitted for simplicity. 
\begin{figure*} 
\centering
\scalebox{0.9}{\includegraphics{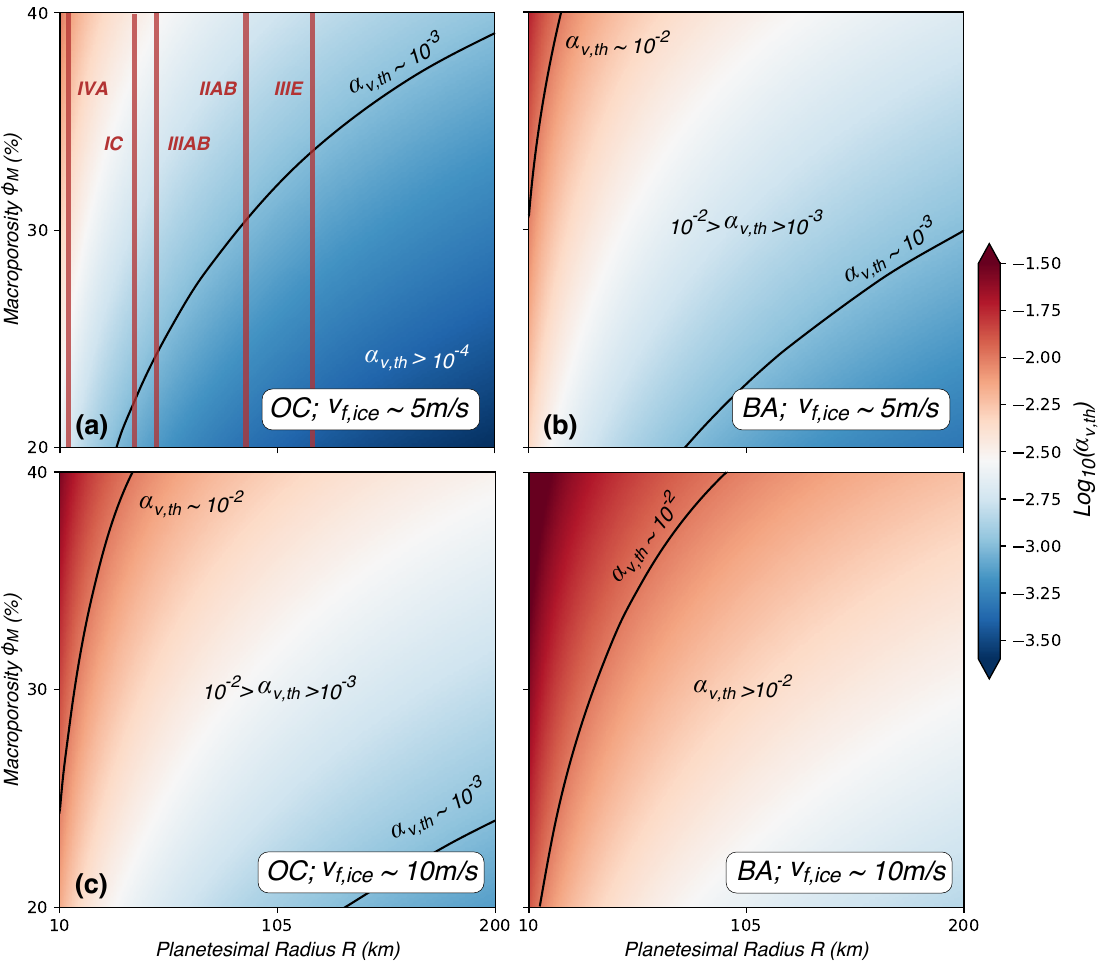}}
\caption{Rasters of the threshold turbulence parameter $\alpha_{\nu,th}$ as a function of planetesimal radius $R$ and macroporosity $\phi_M$ for OC and BA pebbles, two different fragmentation threshold velocities $v_{f,ice}$, and $t_f\simeq 0.5$ Myr after CAIs. Red vertical lines in Fig. 2a correspond to inferred radii of NC IMPBs based on metallographic cooling rates of their respective iron meteorites groups (\textit{e.g.,} IVA) from Qin et al. (2008) (see \textbf{Section 4.3.1} for discussion).}
\label{fig:Figure 5}
\end{figure*}
\begin{figure*} 
\centering
\scalebox{1}{\includegraphics{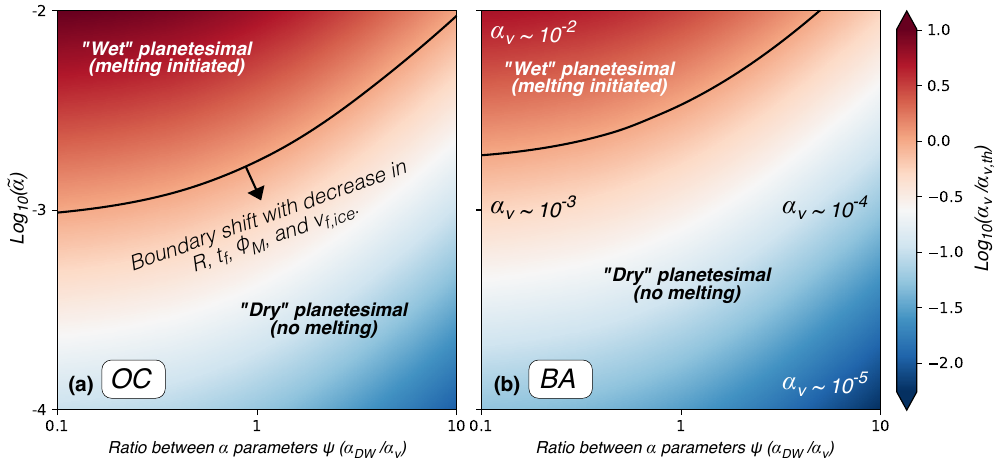}}
\caption{Rasters of $Log_{10}(\alpha_{\nu}/\alpha_{\nu,th})$ across $\tilde{\alpha}-\psi$ parameter space assuming (a) OC and (b) BA pebbles, and fiducial values for planetesimal structural parameters in \textbf{Table 1.} Here, $v_{f,ice}\simeq 5$ m/s. As is clear, no melting of water-ice is expected to occur for quiescent disks, with $\alpha_{\nu}\lesssim 10^{-3}$ (see \textbf{Section 4.3.2} for discussion).}
\label{fig:Figure 6}
\end{figure*}
\subsection{Linking $a_{p,th}$ to $\alpha_{\nu,th}$}
\indent Armed with a ``planetesimal-constrained" $St_{th}$ and a ``disk-constrained" $St_{disk}$, we can proceed to convert $a_{p,th}$ to $\alpha_{\nu,th}$. To do so, and having specified $\psi$, we vary $\alpha_{\nu}$ to seek the value that yields the minimum deviation between $St_{th}$ and $St_{disk}$ at the ice-line. Expressed differently, $\alpha_{\nu,th}$ is that which yields most self-consistency ($St_{th} = St_{disk}$) when our planetesimal and disk models are coupled. In this way, $St$ acts as a bridge between models, yielding $\alpha_{\nu,th}$ by way of numerical brute force computation. In \textbf{Fig. 4}, we present a flowchart of the procedure as described. 
\subsubsection{Turbulence-dominated disk}
\indent Assuming a turbulence dominated disk (\textit{i.e.,} $\psi\simeq0$, $\alpha_{\nu}\simeq \tilde{\alpha}$), we depict in \textbf{Fig. 5} how $\alpha_{\nu,th}$ varies as a function of $R$ and $\phi_M$, for \textbf{(a,c)} OC and \textbf{(b,d)} BA pebbles, $v_{f,ice}\simeq 5$ \textbf{(a,b)} and $10$ \textbf{(c,d) }m/s, and $t_f\simeq 0.5$ Myr. Radii of NC IMPBs inferred from Qin et al. (2008) are marked in \textbf{Fig. 5a} for reference. Several key observations can readily be made. First, in going from OC to BA pebbles (\textit{i.e.,} in increasing the microporosity $\phi_m$), $\alpha_{\nu,th}$ values increase across parameter space, in accord with the decrease in $a_{p,th}$ observed in \textbf{Fig. 3}. Essentially, increased turbulence is required to yield smaller pebbles as the pressure increase in going from macro- to micropores becomes negligible. Comparing upper and lower panels, an increase in $v_{f,ice}$ also leads to an increase in $\alpha_{\nu,th}$, as greater turbulence is required to fragment pebbles of a given size (\textit{i.e.,} $a_{p,th}$). For a planetesimal of $R\simeq 100$ km and $\phi_M\simeq 30\%$, $\alpha_{\nu,th}$ is $\gtrsim 10^{-3}$, suggesting the early SS could not have been quiescent (\textit{i.e.,} $\alpha_{\nu}\simeq 10^{-4}$) for water-ice to have melted in NC IMPBs. Even in the lower right corner of \textbf{Fig. 5a}, at the minimum of the $\alpha_{\nu,th}$ range produced, $\alpha_{\nu,th}$ is $\simeq 3\times10^{-4}$. Recall that $\alpha_{\nu,th}$ corresponds to the onset of melting, at the centermost point in the planetesimal. Disregarding the possibility that a minuscule amount of liquid water could have led to additional melting (\textit{e.g.} via hydrothermal circulation), for several wt.\% liquid water to have been present (Grewal et al., 2024), $\alpha_{\nu}$ must be $> \alpha_{\nu,th}$. The recognition of $\alpha_{\nu,th}$ as a lower limit is also motivated by the potential for (i) pebble growth in clouds (see Section 4.1 above), and (ii) a less massive disk than presently assumed during the early stage of SS formation when IMPBs formed (Lichtenberg et al., 2021). A smaller disk mass corresponds to a smaller surface density and cooler disk, which calls for higher $\alpha_{\nu}$ to produce pebbles $\lesssim$ the threshold size $a_{p,th}$. In each raster, $\alpha_{\nu,th}$ contours ``flatten" towards high $R$, suggesting the formation of larger planetesimals does not alleviate the need for substantial turbulence. 

\subsubsection{Inclusion of MHD winds}
\indent In exploring the behavior of $\alpha_{\nu,th}$ with respect to the structural parameters $R$ and $\phi_M$ in the turbulence-dominated disk, we have obtained a sense of how large $\alpha_{\nu}$ must be to generate pebbles small enough for melting to occur (\textit{i.e.,} $\sim 10^{-3}$). Here, we extend our analysis to disks evolving under MHD winds as well. In particular, we first compute $\alpha_{\nu,th}$ across $\tilde{\alpha}-\psi$ parameter space (once again at the ice-line), bearing in mind each $\tilde{\alpha}$ and $\psi$ pair corresponds to a different disk. We then compare those $\alpha_{\nu,th}$ values to the ``true" $\alpha_{\nu}$ of their respective disks, as calculated with \textbf{Eq. 3}. The region of parameter space wherein $\alpha_{\nu}>\alpha_{\nu,th}$ is that in which liquid water would be produced in the planetesimal interior. For this analysis, we assume $R\simeq 100$ km, $t_f\simeq0.5$ Myr, $\phi_M\simeq 30\%$, and $v_{f,ice}\simeq 5$ m/s. Along with the fiducial values established for other structural parameters (see \textbf{Table 1}), these specifications yields $a_{p,th}\simeq$ 2.2 and 1.5 cm for OC and BA pebbles, respectively. \\
\indent It is worthwhile to note that $\alpha_{\nu,th}$ in our model is a function of $\psi$ but not $\tilde{\alpha}$. This is due to the fact that $\tilde{\alpha}$ does not enter into $\Sigma(r)$, and thus $T(r)$, for our \textit{steady-state} disks, such that it bears no influence on the conversion between $a_{p,th}$ and $St_{th}$, nor the calculation of $St_{disk}$ for any trial $\alpha_{\nu}$ (see \textbf{Fig. 4} and \textbf{Appendix}). Indeed, the utility of $\tilde{\alpha}$ is realized only when the evolution (\textit{i.e.,} decay) of $\Sigma$ with time is considered, in which case $\tilde{\alpha}$ sets the rate thereof (Tabone et al., 2022; Yap \& Batygin, 2024). While the average relative drift velocity between two pebbles of given sizes depends intrinsically on $\tilde{\alpha}$ [$= \alpha_{\nu} (1+\psi)$], which drives the accretionary flow, it is $\alpha_{\nu}$ that is varied to find $\alpha_{\nu,th}$. Thus, for pebble growth limited by relative drift fragmentation, the question underpinning our analysis is: \textit{Given $\psi$, what $\alpha_{\nu}$ yields the $\tilde{\alpha}$ that sets the characteristic pebble radius at the ice-line to $a_{p,th}$?} As it turns out, for both OC and BA pebbles, we find that turbulent fragmentation sets $a_p\simeq a_{p,th}$ across parameter space. That is, $St_{disk}\simeq St_{tf}$. \\
\indent Results from our analysis are depicted in \textbf{Fig. 6}. The key takeaway, as may be intuited from \textbf{Fig. 5}, is that $\alpha_{\nu}$ must be $\gtrsim 10^{-3}$ for melting to occur. For disks that are relatively quiescent, meaning those wind-dominated (high $\psi$) and slow-evolving (low $\tilde{\alpha}$), $\alpha_{\nu}$ is simply too low to yield pebble radii $\lesssim a_{p,th}$. Conversely, disks occupying the upper left corner have $\sim$ an order of magnitude greater than what is required to produce pebbles of radii $\simeq a_{p,th}$. For OC pebbles (\textbf{Fig. 6a}), $St_{th}$ ($\simeq St_{disk}$)  and $\alpha_{\nu,th}$ sit tightly at $\simeq1.65\times 10^{-2}$ and $\simeq 10^{-3.06}$ across parameter space, respectively, with the former (latter) increasing (decreasing) negligibly from low to high $\psi$. For BA pebbles (\textbf{Fig. 6b}), $St_{th}$ and $\alpha_{\nu,th}$ are $\simeq 8.47\times 10^{-3}$ and $\simeq 10^{-2.77}$, respectively, trending with $\psi$ in the same way as for OC pebbles. To summarize, for NC IMPBs to have been ``wet," the (very; $<1$ Myr post-CAIs) early SS must have been considerably turbulent around the ice-line, safely within $10$ AU. \\
\indent To complement \textbf{Fig. 6}, we provide a schematic of our search for $\alpha_{\nu,th}$, given $a_{p,th}$ and $\psi$, in \textbf{Fig. 7}. As is expected with viscous heating, an increase in $\alpha_{\nu}$ leads to that in overall disk temperatures, pushing the ice-line to greater heliocentric distances. The maximal heliocentric distance in the parameter space explored is obtained for a turbulence-dominated disk ($\psi<<1$) with $\alpha_{\nu}\simeq 10^{-2}$, in which the ice-line sits at $\simeq 10$ AU. That said, given that proto-Jupiter formed at the ice-line close to its present-day location at $\sim5.2$ AU (\textit{e.g.,} Raymond et al., 2009; Izidoro et al., 2014), disks characterized by $\alpha_{\nu}$ at the upper end of the range considered (\textit{i.e.,} $\sim 10^{-2}$) appear to have ice-lines that are too far from the Sun. While it is tempting to suggests that $\alpha_{\nu}$ is thus well constrained to be just in excess of $10^{-3}$, it is important to keep in mind that our disk model is not anchored to CAI formation (the conventional SS time zero), nor are its assumed parameters (\textit{e.g.,} initial disk mass; see \textbf{Appendix}) undoubtedly representative of the early SS. Stated differently, $\Sigma(r)$ for our disks does not necessarily correspond to a post-CAI time of $t_f\simeq 0.5$ Myr, as assumed for NC IMPBs in our analysis. Inward drift of the ice-line accompanies the decay in $\Sigma$ over time. As such, if the $\Sigma(r)$ profiles adopted reflect a pre-CAI disk, an ice-line close to $\sim5$ AU by $t_f\simeq 0.5$ Myr may be conceivable despite high $\alpha_{\nu}$ (note that a decrease in $\Sigma$ ultimately leads to an increase in $\alpha_{\nu,th}$ for a given $a_{p,th}$; see equations for $St(a_p)$ and $St_{tf}$ in the \textbf{Appendix}). We emphasize that these complications/degeneracies do not influence the first-order result of our work: the early SS could not have been quiescent ($\alpha_{\nu}\lesssim$ few times $10^{-4}$) if liquid water was abundant in the oldest NC planetesimals (see also \textbf{Fig. 5}). 
\begin{figure} 
\scalebox{1.03}{\includegraphics{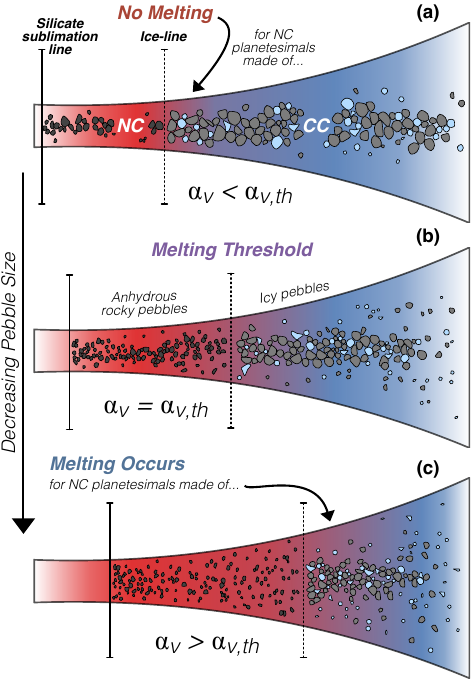}}
\caption{Illustration of a SS disk with $\alpha_{\nu}$ (a) $<\alpha_{\nu,th}$, (b) $=\alpha_{\nu,th}$, and (c) $> \alpha_{\nu,th}$. Melting of water-ice in the interior of the planetesimal formed by the agglomeration of icy pebbles beyond the ice-line is achievable in (b) and (c). As $\alpha_{\nu}$ increases (for a given $\psi$), pebble sizes across the disks decrease, and the ice-line sits farther from the Sun (see Section \textbf{4.3.2} for discussion). }
\label{fig:Figure 7}
\end{figure}

\section{Discussion}
\subsection{Comparison to empirical constraints on turbulence}
\indent Our estimates for $\alpha_{\nu}$ in the early SS derive from a confluence of meteoritics, fluid dynamics, and disk theory. Here, we consider how these estimates fare with empirical constraints on turbulence from telescopic investigations of protoplanetary disks, primarily through ALMA (see Rosotti, 2023 for a review). Such investigations may be categorized broadly into two groups: those measuring turbulence ``directly" via turbulent broadening of spectral lines, typically in the infrared (IR) or sub-mm (\textit{e.g.,} Carr et al., 2004; Flaherty et al., 2020), and those that do so ``indirectly" by assessing the physical consequences of turbulence, such as the scale height of the dust sub-disk (\textit{e.g.,} Villenave et al., 2020) and the rate of viscous spreading from angular momentum redistribution (\textit{e.g.,} Trapman et al., 2020). Excluding direct detection in the IR, which probes $\alpha_{\nu}$ at small heliocentric distances ($r\sim 0.1$ AU), all methods do so at $r$ in the tens to hundreds of AU. As such, no direct comparison can be made to our estimates for $\alpha_{\nu}$ around the ice-line (typically $2\lesssim r\lesssim 10$ AU; \textit{e.g.,} Yap \& Batygin, 2024).  It remains worthwhile, nonetheless, to gauge the range of $\alpha_{\nu}$ measured so as to evaluate the soundness of our results, as least to first order. \\
\indent The characteristic velocity scale for turbulent eddies, assuming the turnover frequency of the largest eddies is comparable to the Keplerian orbital frequency, is given by $\sqrt{\alpha_{\nu}}$ times the isothermal sound speed $c_s$ (\textit{e.g.,} Sengupta et al., 2024). Observations of IR line broadening indicate turbulent velocities comparable to $c_s$ at the disk inner edge, corresponding to $\alpha_{\nu}\gtrsim 10^{-2}$. Such high values are consistent with expectations of high ionization in proximity to the star, and especially at upper disk layers, where ideal MHD can be achieved (\textit{e.g.,} Armitage, 2020). Far from the star, analyses of sub-mm line broadening and indirect methods yield an array of $\alpha_{\nu}$ values suggestive of disk diversity, ranging between $10^{-4}$ to $10^{-2}$. Sub-mm observations yield $\alpha_{\nu}$ values towards the upper end of this range, though results are far from clustered. For instance, using CO and DCO$^+$ as tracers, Flaherty et al. (2017) obtained an upper limit of $\alpha_{\nu}\simeq 2.5\times 10^{-3}$ for HD163296. This limit is $\sim$ an order of magnitude smaller than the $\alpha_{\nu}$ measured for DM Tau using CO alone, at $\simeq 6.3 \times 10^{-2}$ (Flaherty et al., 2020). \\
\indent Indirect methods have been more extensively applied, and yield $\alpha_{\nu}$ values at the lower end of the said range. The most prevalent of such methods involves the measurement of the degree of dust settling (\textit{i.e.,} its vertical extent). It is well established that pebbles settle into a solid sub-disk about the disk midplane, the scale height of which (denoted $h_{d}$) depends on their $St$, setting the extent to which turbulent diffusion can compete against gravitational settling. The expression for $h_d$ takes the form (Dubrulle et al., 1995)
\begin{equation}
\frac{h_d}{h} = \left(1+\frac{St}{\alpha_{\nu}}\right)^{-1/2},
\end{equation}
where $h$ is the hydrostatic scale height of the disk gas. Measurements of $h_d/h$ from ALMA's Disk Substructures at High Angular Resolution Project (DSHARP) yield values of $\alpha_{\nu}/St$  ranging between few times $10^{-2}$ to few times $0.1$. (Dullemond et al., 2018; Rosotti et al., 2020). Assuming $St\simeq 10^{-2}$ to $0.1$ (and bearing in mind this source of degeneracy), $\alpha_{\nu}$ is centered around $10^{-3}$. Other indirect methods (see \textbf{Table 3} in Rosotti, 2023) typically yield $\alpha_{\nu}\lesssim 10^{-3}$. \\
\indent Our region of interest for NC IMPBs is bracketed by disk domains probed with astronomy. With $\alpha_{\nu}\gtrsim 10^{-2}$ measured close to the star, and $\alpha_{\nu}\lesssim 10^{-3}$ measured far from it, respectively, telescopic investigations at the population level are concordant with our estimated lower limit of $\alpha_{\nu}\gtrsim 10^{-3}$ for the early SS. 

\subsection{Implications for terrestrial planetary accretion}
\indent Planetesimals, having formed, can proceed to grow by either (i) mutual collisions aided by gravitational focusing (\textit{e.g.,} Safronov, 1972; Greenberg et al., 1978; Kokubo \& Ida, 1998) or (ii) pebble accretion (\textit{e.g.,} Ormel \& Klahr, 2010; Lambrechts \& Johansen, 2012). The dominant mechanism depends largely on the efficiency of the latter, itself dependent on the scale height of the dust sub-disk from above (\textbf{Eq. 5}; Ormel, 2017). Pebbles with low $St$ are tightly coupled to the gas, and so remain vertically dispersed about the midplane. In this case, pebble accretion operates in the so-called ``3D" regime. Conversely, pebbles with high $St$ settle readily towards the midplane, allowing planetesimals to partake in the more efficient ``2D" regime. For a given $St$, it is clear that higher $\alpha_{\nu}$ leads to a greater inclination for 3D pebble accretion. \\
\indent Recently, Yap \& Batygin (2024) argued that pebble accretion can only dominate rocky planet growth within the ice-line in extremely quiescent disks ($\alpha_{\nu}\sim 10^{-5}$), that is, those characterized by low $\tilde{\alpha}$ and high $\psi$ (see \textbf{Section 4}; note the flipped inequality sign in Fig. 12a of their paper). Our results constraining $\alpha_{\nu}$ around the formation region of rocky NC planets in the early SS to at least $\gtrsim$ few times $10^{-4}$ are thus in favor of their growth by planetesimal collisions (assuming $\alpha_\nu$ does not decay appreciably, and rapidly, across the $\sim 4$ to $5$ Myr circumsolar disk lifetime; Weiss et al., 2021). This is consistent with nucleosynthetic isotope anomalies suggesting Earth (and Mars) did not accrete substantial CC materials as inward drifting pebbles (\textit{e.g.,} Burkhardt et al., 2021; Kleine et al., 2023; Morbidelli et al., 2025). In fact, reconstructions of Earth's accretion history using elements (i) exhibiting such anomalies and (ii) characterized by varying siderophilic affinities (or ``mantle memory" of accretion; \textit{e.g.,} Mo, Fe, Cr, Ni) suggest our planet accreted mostly NC materials in its initial $\sim60$ to $80\%$ of growth (Dauphas et al., 2017; Dauphas et al., 2024). This first-order finding is corroborated by studies focused on the Earth's volatile inventory, pointing to its origin from ``dry," differentiated planetesimals deep within the inner SS (and NC reservoir), perhaps close to the silicate sublimation line at $T\simeq 1400$ K (Morbidelli et al., 2020; Sossi et al., 2022; Tissot et al., 2022; Liu et al., 2023). CC planetesimals contributing to the final $\sim 20\%$ of Earth's mass are suggested to have been dynamically introduced into the inner SS following disk dissipation at $\simeq 4$ Myr post-CAIs (Dauphas \& Chaussidon, 2011; Wang et al., 2017; Borlina et al., 2022; Kleine et al., 2024; Nimmo et al., 2024). \\
\indent The ``dry" accretion of the early Earth naturally begs the question as to how it acquired its present-day abundance of water, an ongoing subject of debate (\textit{e.g.,} O'Brien et al., 2018). While NC IMPBs may have been ``wet," explaining the high oxidation states inferred from their cores, it is unlikely that they contributed a significant portion of Earth's water. This follows from the low water contents measured in NC (and CC) achondrites, testifying to substantial water loss during differentiation (Newcombe et al., 2023), and supporting the need for either late accretion of CC chondrites, cometary material, and/or interstellar ices (McCubbin et al., 2019) to Earth, and/or extensive H$_2$ ingassing (\textit{e.g.,} Olson \& Sharp, 2018) during the lifetime of the SS disk. The likely ``dry" nature of NC IMPBs following differentiation does not necessarily imply that they reflect the primary building blocks of the Earth. As hinted above, Earth may owe its accretion to early planetesimals constructed from anhydrous dust piled up at the silicate line (see framework of planetesimal ``ring" from Batygin \& Morbidelli, 2023), isotopically akin to, and yet chemically distinct from materials formed at the ice-line (Morbidelli et al., 2020; Tissot et al., 2022).\\
\indent As a final point, note that pebble accretion can, and likely does, dominate planetary growth in the outer disk (beyond the ice-line) where characteristic pebble $St$ (equivalent to $St_{disk}$ in our model) can be $\sim$ an order of magnitude larger than they are in the inner disk, resolving timescale conflicts associated with giant planet cores (Lambrechts \& Johansen, 2012; Yap \& Batygin, 2024). 

\section{Concluding Remarks}
\indent Motivated by emerging evidence for liquid water in NC IMPBs, we have utilized a model for a degassing icy planetesimal to find a threshold characteristic pebble radius below which melting of water-ice occurs. Adopting a disk model, we translated the said radius to a lower limit on the degree of turbulence around the ice-line in the early SS. Key to this translation is the Stokes number, linking pebbles composing the planetesimal to their fragmentation-limited growth in the circumsolar disk. \\
\indent Major gaps in our understanding of SS history stem from the mutually orthogonal approaches with which researchers in different disciplines have investigated the early SS. Our work is an attempt to remedy part of this scholastic divide, uniting cosmochemistry and theoretical astrophysics in constraining early SS turbulence. 

\section*{Appendix I: \\ The Icy Planetesimal}

\indent Central to this study is the ``two-scale" model for a degassing icy planetesimal described in Section 2 and inspired by Zhang (2023). As a brief overview, the planetesimal of radius \textit{R} forms at $t_{f}$ after CAI condensation (\textit{i.e.,} SS time zero), and consists of rocky and water-ice pebbles of characteristic size/radius $a_{p}$, the packing of which yields a macroporosity $\phi_{M}$. The rocky pebbles consists of silicate and water-ice grains of characteristic size $a_{g}$, the packing of which yields a microporosity $\phi_{m}$. The rocky pebbles, silicate grains, and water-ice have densities $\rho_r$, $\rho_s$, and $\rho_i$, respectively. For self-consistency in using present-day, anhydrous, materials (\textit{e.g.,} chondrites) as pebble analogs, we relate their microporosity $\phi_X$ ($X$ denoting the analog: OC or BA; see \textbf{Section 2.1}) to $\phi_m$ with the ice-to-silicate mass ratio (at time $t_{f}$), denoted $f$ (see Table 1 for the compilation of key parameters describing the planetesimal structure). Below, we first relate $\phi_m$, $\phi_X$, and $f$ before providing a full derivation of the (macro- and micro-) pore pressures required to drive vapor loss.

\subsection*{Relating Microporosities}
\indent At the pebble-scale, the specification of $\rho_{s}$, $\rho_{i}$, and $\phi_X$ enforces a maximum $f$, corresponding to all interstices between silicate grains being filled with ice. This value, denoted $f_{max}$, takes the form
\[
f_{max} = \frac{\rho_i \phi_X}{\rho_s (1-\phi_X)}.
\]
The microporosity $\phi_m$ is given by
\[
\phi_m = \phi_X - \frac{f(1-\phi_X)\rho_s}{\rho_i}.
\]
In our model, we set $f\simeq 0.5 f_{max}$, such that  $\phi_m \simeq \phi_X/2$ (\textit{i.e.,} half the available space for ice is filled). The rocky pebble density $\rho_r$ is given by $\rho_{X} (1+f)$.

\subsection*{Deriving Pore Pressures}
\indent At the pebble-scale, degassing is driven by the gradient between micro- and macropore pressures, $P_m$ and $P_M$, respectively. Likewise, the gradient between macropore pressures and the disk pressure at the planetesimal surface $P_{D}$ drives degassing at the planetesimal-scale. The key parameter relating $P_m$ to $P_M$, and $P_M$ to $P_d$, is the permeability $\kappa$. Intuition suggests $\kappa$ ought to depend on the characteristic pebble/grain size and the porosity, and indeed it does. To parameterize $\kappa$, we adopt the Kozeny-Carman equation, which assumes the porous medium across which a fluid flows can be modeled as a series of conduits with constant cross-sectional area but differing and complex shapes (Dullien, 2012). The micropermeability $\kappa_m$, then, is given by
\[
\kappa_m \simeq \frac{a_g^{2}\phi_m^{3}}{45(1-\phi_m)^{2}}.
\]
Similarly, the macropermeability is given by
\[
\kappa_M \simeq \frac{a_p^{2}\phi_M^{3}}{45(1-\phi_M)^{2}}.
\]
It is useful to define the characteristic micropore size $a_m$ as well:
\[
a_m \simeq \frac{2 a_g\phi_m}{3(1-\phi_m)}
\]
as its relation to the vapor mean free path $\lambda$ will determine the regime in which flow operates within the rocky pebbles. From the kinetic theory of gases, $\lambda$ can be expressed as
\[
\lambda \simeq \frac{\nu}{P_m}\sqrt{\frac{\pi k_B T}{2\mu}},
\]
where $\nu \simeq 10^{-5}$ Pa$\cdot$ s is the vapor viscosity at $T\simeq 273$K (Sengers \& Kamgar-Parsi, 1984), $k_B = 1.38\times 10^{-23}$ m$^{2}$kg/s$^{2}$K is the Boltzmann constant, and $\mu\simeq 3\times 10^{-26}$ kg is the mean molecular mass of H$_2$O. \\
\indent Let us first derive the relationship between $P_m$ and $P_M$. The total mass flux, accounting for the transition between gas transport in the Poiseuille ($\lambda/a_m << 1$) and Knudsen  ($\lambda/a_m >> 1$) regimes is given by (Klinkenberge, 1941)
\[
\vec{q_m} = -\frac{P_m\mu\kappa_m}{\nu k_B T}\left(1+\frac{4\lambda}{a_m}\right)\vec{\nabla}P_m.
\]
Note that the volume flux $\vec{q_v}$ can be obtained simply by dividing $\vec{q_m}$ by the gas density $\rho_g = P_m\mu/k_B T$, as given by the ideal gas law. For $\lambda<<a_m$, Darcy's Law describing Poiseuille flow is recovered:
\[
\vec{q_{m,Poi}} = -\frac{P_m\mu\kappa_m}{\nu k_B T}\vec{\nabla}P_m.
\]
For $\lambda>>a_m$, the equation reduces instead to that for Knudsen flow:
\[
\vec{q_{m,Knud}} = -\frac{4\lambda P_m\mu\kappa_m}{\nu k_B Ta_m}\vec{\nabla}P_m.
\]
(For gas flow through the macropores between pebbles later on, Poiseuille flow dominates due to the large macropore size). Substituting $\lambda$ and $a_m$ into $\vec{q_m}$, and assuming spherical symmetry, we obtain
\[
q_m = -\frac{P_m\mu\kappa_m}{\nu k_B T}\left(1+\frac{6\nu(1-\phi_m)}{a_g\phi_m P_m}\sqrt{\frac{\pi k_B T}{2\mu}} \right)\frac{dP_m}{dr_m},
\]
$r_m$ being the distance from the center of a rocky pebble. \\
\indent With an expression for the outward flux of water vapor defined, let us turn our attention to vapor generation by $^{26}$Al decay. We first derive the rate of energy production per unit mass from such decay, which we denote $\theta$. As $^{26}$Al is virtually extinct today, the present-day abundance of Al reflects solely that of $^{27}$Al, the only stable and naturally-occurring isotope of Al. The mass fraction of $^{27}$Al thus calculated is [Al]$_{today}\simeq 8.68\times 10^{-3}$, and the number of $^{27}$Al atoms $f_{27} \simeq$ [Al]$_{today}/(27m_p)$, where $m_p\simeq 1.67\times 10^{-27}$kg is the proton mass. The atomic abundance of $^{26}$Al at some time $t$ following the CAI formation $n_{26}(t)$ is simply $f_{27}\times $($^{26}$Al/$^{27}$Al)$_t$, where 
\[
\left(\frac{^{26}Al}{^{27}Al}\right)_t = \left(\frac{^{26}Al}{^{27}Al}\right)_{CAI} e^{-\lambda_{26}t}
\]
from the decay equation. The canonical CAI initial $^{26}$Al/$^{27}$Al is $\simeq 5.2\times 10^{-5}$ (\textit{i.e.,} MacPherson et al., 1995; Jacobsen et al., 2008), and the $^{26}$Al decay constant $\lambda_{26}\simeq 3.06\times 10^{-14}$s$^{-1}$ (half-life $\simeq 0.717$ Myr). Each decay produces an energy $E_d\simeq 3.3 $ Mev $\simeq 5.3\times 10^{-13}$J. Thus, the energy production rate per unit mass at time $t$ is
\[
\begin{split}
\theta &= E_d\frac{dn_{26}}{dt} = E_d\lambda_{26}n_{26}(t)\\
&= \frac{E_d\lambda_{26}[Al]_{today}(^{26}Al/^{27}Al)_{CAI}e^{-\lambda_{26}t}}{27m_p}.
\end{split}
\]
\indent The energy production rate within a sphere of radius $r_m$ from the center of a rocky pebble is expressed as
\[
Q_m = \frac{4}{3}\pi r_m^{3}(1-\phi_X)\rho_{s}\theta,
\]
where the factor $1-\phi_X$ is the fraction of silicates within the sphere. The vapor production rate within the sphere is simply $Q_m/\Delta H$, where $\Delta H\simeq 2.8 \times 10^{6}$ J/kg is the H$_2$O latent heat of sublimation (Feistel \& Wagner, 2007). Now, we impose that vapor production is equal to vapor loss, or
\[
q_m = \frac{Q_m}{4\pi r_m^{2}\Delta H}.
\]
This condition yields not $P_m$ at the onset of sublimation, but that which is needed to expel the vapor produced from the pebble. If $4\pi r_m^{2} q_m < Q_m/\Delta H$, pressure buildup will proceed until $4\pi r_m^{2} q_m = Q_m/\Delta H$. We simply ask whether $P_m$ at that steady-state point is above or below the triple point pressure of 6 mbar. \\
\indent Equating the two expressions for $q_m$ derived, and solving the differential equation with the boundary conditions $P_m(r_m=a_p)=P_M(r_M)$, where $r_M$ is the radial distance from the center of the planetesimal to the location of the rocky pebble, we arrive at a quadratic equation for $P_m$:
\[
\begin{split}
&\frac{\mu\kappa_m}{2\nu k_B T}[P_m^{2}-P_M^{2}(r_M)] + \frac{6\kappa_m(1-\phi_m)}{a_g\phi_m}\sqrt{\frac{\pi\mu}{2k_B T}}...\\ &...[P_m - P_M(r_M)]
- \frac{(1-\phi_X)\rho_{s}\theta}{6\Delta H}(a_p^{2}-r_m^{2}) = 0.
\end{split}
\]
This can be solved once we obtain an expression for $P_M(r_M)$ below. For clarity, the terms $p_i$ corresponding to the second-, first-, and zeroth-order powers of $P_m$ are:
\[
\begin{split}
&p_2 \simeq \frac{\mu a_g^{2}\phi_m^{3}}{90\nu k_B T(1-\phi_m)^{2}}\\
& p_1 \simeq \frac{2a_g\phi_m^{2}}{15(1-\phi_m)}\sqrt{\frac{\pi\mu}{2k_B T}}\\
p_0 \simeq &-\left[p_2 P_M^{2} + p_1 P_M + \frac{(1-\phi_X)\rho_{s}\theta (a_p^{2}-r_m^{2})}{6\Delta H}\right],
\end{split}
\]
where we have expressed $\kappa_m$ in terms of $a_g$ and $\phi_m$. \\
\indent Now, let us consider the macropore pressure $P_M$. As mentioned above, gas transport between the pebbles takes place in the Poiseuille regime. As such, the mass flux $\vec{q_M}$ takes the form 
\[
\vec{q_M} = -\frac{P_M\mu\kappa_M}{\nu k_B T}\vec{\nabla}P_M.
\]
Within a sphere of radius $r_M$, the energy production rate is given by
\[
Q_M = \frac{4}{3}\pi r_M^{3} (1-2\phi_M)(1-\phi_X)\rho_{s}\theta,
\]
where we have assumed that $\simeq$ half the space between rocky pebbles is occupied by ice, and the factor $1-\phi_X$ accounts for the fact that only the silicate portion of rocky pebbles host $^{26}$Al. Equating the vapor production in the sphere $Q_M/\Delta H$ to the vapor loss rate $4\pi r_M^{2}q_M$, we obtain
\[
q_M = \frac{r_M(1-2\phi_M)(1-\phi_X)\rho_{s}\theta}{3\Delta H}.
\]
Finally, equating the two expressions for $q_M$ derived, assuming (once again) spherical symmetry, and solving the resulting differential equation with the boundary condition $P_M(r_M=R) = P_o$, we obtain
\[
P_M \simeq \sqrt{\frac{15(1-2\phi_M)(1-\phi_X)\nu k_B T \rho_{s}\theta(R^{2}-r_M^{2})}{\mu a_p^{2} \phi_M^{3}(1-\phi_M)^{-2}\Delta H } + P_o^{2}}.
\]
\indent With $P_M$ defined, it is now possible to solve the quadratic equation from above for $P_m$. In our model, we set $r_m= r_M = 0$, meaning we evaluate the micropore pressure at the very heart of the planetesimal (see \textbf{Section 2.2}). In the main text, we denote $P_m(r_m=r=0)$ as $P_{m,c}$ for convenience. If $P_{m,c}$ exceeds 6 mbar, melting occurs. Only under the condition that $P_m(r_m=r=0)\lesssim 6$ mbar is it appropriate (\textit{i.e.,} self-consistent) to vary $r_M$ to find the pressure within a pebble at some location above the planetesimal center. 

\section*{Appendix II: \\The Disk \& Stokes Number}

\indent Translating the threshold characteristic pebble radius $a_{p,th}$ to a constraint on early SS turbulence (\textit{i.e.,} $\alpha_{\nu,th}$) requires a disk model with which we can (i) convert $a_{p,th}$ to a threshold Stokes number $St_{th}$, and (ii) compute the characteristic Stokes number of disk pebbles as defined by collisional fragmentation $St_{disk}$ (\textit{i.e.,} the smaller between $St_{tf}$ \& $St_{df}$; the maximal $St$ permitted by turbulent and relative drift fragmentation, respectively). Recall that $\alpha_{\nu,th}$ is obtained through comparison of $St_{disk}$ to $St_{th}$ (see \textbf{Section 4} \& \textbf{Fig. 4}). Here, we provide an outline of the disk model adopted (see Yap \& Batygin, 2024 for an in-depth description), the equations for $St_{tf}$ and $St_{df}$, as well as the relationship between $a_{p,th}$ and $St_{th}$ in the Epstein drag regime. 

\subsection*{Disk Model}
\indent We assume a steady-state disk subject to angular momentum transport by both turbulence and magnetohydrodynamic (MHD) winds (Tabone et al., 2022; Yap \& Batygin, 2024). The combined torque exerted on the disk by these two mechanisms is quantified by $\tilde{\alpha}$, constituting a sum of the Shakura-Sunyaev turbulence parameter $\alpha_{\nu}$ and an analogous parameter $\alpha_{DW}$, representing the contribution from winds: 
\[
\tilde{\alpha} = \alpha_{\nu} + \alpha_{DW}.
\]
The relative strength between turbulence and winds is quantified the ratio of $\alpha_{DW}$ to $\alpha_{\nu}$, denoted $\psi$:
\[
\psi = \alpha_{DW}/\alpha_{\nu}.
\]
\indent Assuming a disk midplane temperature scaling with heliocentric distance $r$ as $T\sim r^{\gamma -3/2}$, the disk surface density $\Sigma(r)$ takes the form
\[
\Sigma(r) = \Sigma_{c}\left(\frac{r}{r_c}\right)^{\xi-\gamma}exp\left[-\left(\frac{r}{r_c}\right)^{2-\gamma}\right],
\]
inspired by the analytical solution for the $\Sigma$ evolution under constant $\alpha_{\nu}$ by Lynden-Bell \& Pringle (1974). Here, $\Sigma_c$ is the characteristic surface density, and $r_c$ the characteristic disk outer radius at which $\Sigma \simeq $ $\Sigma_c/e$. The former can be expressed in terms of the disk total mass $M_{disk}$ as
\[
\Sigma_c \simeq \beta\frac{M_{disk}}{2\pi r_c^{2}},
\]
where $\beta$ is a factor of order unity introduced to ensure $\int^\infty_{0.1AU} 2\pi r \Sigma(r) dr$ retrieves $M_{disk}\simeq 0.05 M_{\odot}$ ($M_{\odot}\simeq 2\times 10^{30}$ kg). This choice of $M_{disk}$ yields $\Sigma(r=1 AU)$ close to that in the \textit{Minimum Mass Solar Nebula} model of Hayashi (1981) (\textit{i.e.,} $\sim 2 \times 10^{4}$ kg/m$^{2}$) for an intermediate $\alpha_{\nu}$ value of $\sim 10^{-3}$. We set $r_c \simeq 12$ AU, bearing in mind that its exact value is insignificant so long as it lies away from the ice-line ($< 10$ AU). Note that $0.1$ AU is taken as the disk inner edge, corresponding to the magnetospheric truncation radius. The term $\xi$ is the so-called ``mass ejection index" (Tabone et al., 2022), serving to flatten the $\Sigma$ profile in reflecting mass extraction from winds. It is given by
\[
\xi = \frac{\psi+1}{4}\left(\sqrt{1+\frac{4\psi}{(\lambda-1)(\psi+1)^2}}-1\right),
\]
where $\lambda\simeq 3.5$ is the magnetic lever arm parameter, a measure of the angular momentum per unit mass extracted by disk wind streamlines (Blandford \& Payne, 1982). Accordingly, $\gamma$ is the power law index of $\Sigma(r)$ in disks dominated by turbulence (\textit{i.e.,} $\psi\simeq 0$ $\Rightarrow$ $\xi\simeq 0$), and at small $r$ (before the exponential decay term takes hold). 
\indent The disk is assumed to be heated by viscous shear between disk annuli alone, as stellar irradiation is negligible in the region of interest $r << r_c$ (Chambers, 2009; Yap \& Batygin, 2024). Its relevance increases, nonetheless, with lower $\alpha_{\nu}$ and larger $r_c$. The midplane temperature of such a disk $T$ is established via equilibrium between heat generation and radiative loss, expressed as (Armitage, 2020)
\[
\sigma T^{4} \simeq \frac{3}{4}\tau F_{visc},
\]
where $\sigma = 5.67\times10^{-8}$ W/m$^2$K$^4$ is the Stefan-Boltzmann constant, $\tau$ the optical depth, and $F_{visc}$ the heating rate per unit area. The latter is given by (Nakamoto $\&$ Nagakawa, 1994) 
\[
F_{visc} = \frac{1}{2}\Sigma\nu\left(r\frac{\partial \Omega_K}{\partial r}\right)^2 = \frac{9}{8}\Sigma\alpha_\nu c_s^{2}\Omega_K,
\]
where the turbulent viscosity $\nu = \alpha_\nu c_s^2/\Omega_K$ (Shakura \& Sunyaev, 1973), with $c_s = \sqrt{k_B T/\mu'}$ being the isothermal sound speed and $\Omega_K =\sqrt{GM_{\odot}/r^3}$ the Keplerian angular velocity. The mean molecular mass of the H$_2$-He disk gas mixture $\mu'\simeq 2.4 m_p$, the gravitational constant $G\simeq 6.67\times 10^{-11}$ m$^3$/kgs$^2$, and the solar mass $M_{\odot}\simeq 2\times 10^{30}$ kg. The optical depth takes the form (Bitsch et al., 2014)
\[
\tau = \frac{1}{2}f_d k_d \Sigma,
\]
where the dust-to-gas ratio (\textit{i.e.,} metallicity) $f_d \simeq 0.01$, and the dust opacity $k_d\simeq 30$ m$^2$/kg. Altogether, the above equations yield
\[
T_{disk} = \left(\frac{27f_d k_d \alpha_\nu k_b \Omega_K \Sigma^2}{64\sigma \mu'}\right)^{1/3}.
\]
\indent The coupled scaling of the midplane temperature ($\sim r^{\gamma - 3/2}$) and surface density ($\sim r^{\xi - \gamma}$) profiles is enforced by the condition that, for $r<r_c$, the disk is defined by a constant stellar mass accretion rate $\dot{M_\odot}\sim \Sigma \nu \sim r^{0}$ (note $\xi$ is constant) under the Shakura-Sunyaev $\alpha$-prescription for viscosity. The self-consistent choice for $\gamma$  yields a $\Sigma$ profile which, upon substitution into the above equation for $T$, leads to a $T$ profile with a power law index ($\gamma -3/2$) concordant with that $\gamma$. We find that $\gamma\simeq 0.6$ results in power law indices for $T$ consistent to within $5\%$ across the range of $\alpha_\nu$ explored. Moreover, with $\gamma$ determined, we find that $\beta\simeq1.45$ yields $\Sigma$ profiles that reproduce $M_{disk}$ to within $3\%$. 

\subsection*{The Characteristic Stokes Number:\\ Disk-constrained}
\indent As discussed in \textbf{Section 4}, the characteristic $St$ of disk pebbles (\textit{i.e.,} $St_{disk}$) is given by the smaller of the maximal $St$ permitted by turbulent and relative drift fragmentation (\textit{i.e.,} $St_{tf}$ and $St_{df}$, respectively). The former is expressed through the equivalence between the average relative velocity of similar-sized pebbles in turbulent motion $\Delta v_t = \sqrt{3\alpha_\nu St}c_s$ and the fragmentation threshold of those pebbles $v_f$ (Ormel \& Cuzzi, 2007):
\[
St_{tf} = \frac{1}{3}\frac{v_f^2}{\alpha_\nu c_s^2}.
\]
Recall that $v_f$ for the icy pebbles in composing the planetesimal in our model is assumed to be $\simeq 5$ or $10$ m/s (see \textbf{Section 4.2} \& \textbf{Fig. 5}). \\
\indent To obtain $St_{df}$, we equate some characteristic relative drift velocity between similar-sized pebbles $\Delta v_r$ to $v_f$. The radial drift velocity of pebbles subject to drag in protoplanetary disks takes the form (Nakagawa et al., 1986, Armitage, 2020)
\[
v_r = \frac{St^{-1}v_{r,gas} + |\epsilon|(c_s^2/v_K)}{St + St^{-1}},
\]
where $v_{r,gas} = (3/2)(2-\gamma)^2 \tilde{\alpha} c_s^2 v_K^{-1}$ (Tabone et al., 2022) is the radial drift velocity of disk gas, $v_K = r\Omega_K$ is the Keplerian orbital velocity, and $|\epsilon| = 3.5|0.5+\gamma/3 -\xi/3|$ the power law index of the disk's midplane pressure profile. Assuming the smaller of the colliding pebbles has $St_2 = N St_1$, where $N\simeq0.5$ (Birnstiel et al., 2012), $\Delta v_r$ can be expressed as
\[
\begin{split}
\Delta v_{r} = \frac{c_s^{2}}{v_{K}}&\left(\frac{3[\alpha_\nu/(1+\psi)](2-\gamma)^{2}/2 + |\epsilon|St_{1}}{St_{1}^{2}+1}\right.\\&\ \ \ \ \ \ \ \ \left.- \frac{3[\alpha_\nu/(1+\psi)](2-\gamma)^{2}/2 + |\epsilon|NSt_{1}}{N^{2}St_{1}^{2}+1} \right).
\end{split}
\]
Setting $\Delta v_{r} = v_{f}$ and rearranging for $St_{1}$ (\textit{i.e.,} $St_{df}$) yields a quartic equation with coefficients given by
\[
\begin{split}
p_{4} = \frac{v_{f}v_{K}N^{2}}{c_s^{2}} &;\ p_{3} = |\epsilon|(N-N^{2})\\
p_{2} = \frac{v_{f}v_{K}(N^{2}+1)}{c_s^{2}} &+ \frac{3[\alpha_\nu/(1+\psi)](2-\gamma)^{2}(1-N^{2})}{2}\\
p_{1} = |\epsilon|(N-1)&;\ p_{0} = \frac{v_{f}v_{K}}{c_s^{2}}
\end{split}
\]
where $p_{i}$ corresponds to the $i$th-order term in $St_{df}$. Note that $\Delta v_r$ is intrinsically dependent on $\tilde{\alpha}$, but expressed in terms of $\alpha_{\nu}$ and $\psi$ above. This is done to emphasize that $\alpha_{\nu}$ is the parameter of interest that is varied in the computational search for $\alpha_{\nu,th}$ corresponding to some $a_{p,th}$. Accordingly, in \textbf{Section 4.3.2}, disks that share the same $\psi$ but different $\tilde{\alpha}$ (\textit{i.e.,} those sharing a column in the rasters of \textbf{Fig. 6}) have the same $\alpha_{\nu,th}$, despite different actual $\alpha_{\nu}$ as computed by \textbf{Eq. 3}, given here for completeness:
\[
\alpha_{\nu} = \tilde{\alpha}/(1+\psi).
\]
As a reminder (\textbf{Eq. 4}),
\[
St_{disk} = min[St_{tf},St_{df}].
\]
\subsection*{The Characteristic Stokes Number:\\ Planetesimal-Constrained}
\indent  Having obtained a characteristic pebble radius $a_{p,th}$ (via specification of planetesimal structural parameters and post-CAI formation time) we convert it to $St_{th}$ via
\[
St_{th} = \frac{a_{p,th}\pi\rho_r}{2\Sigma},
\]
valid in the Epstein regime, in which $a_{p,th}\lesssim 9/4$ times the mean free path of disk gas (largely H$_2$) molecules $\lambda_{mfp}$, where
\[
\lambda_{mfp} \simeq \frac{\sqrt{2\pi}c_s \mu'}{\Sigma\Omega_K\sigma_{mol}},
\]
with $\sigma_{mol}\simeq 2\times 10^{-19}$ m$^2$ the cross-section for H$_2$ mutual collisions (Armitage, 2020). For every $\alpha_{\nu}$ tested in searching for $\alpha_{\nu,th}$ (with which $St_{th}\simeq St_{disk}$), we calculated $\lambda_{mfp}$ to ensure pebbles with radius $a_{p,th}$ were in the Epstein regime before proceeding to calculate $St_{th}$. Disk parameters in both $St_{th}$ and $\lambda_{mfp}$ were evaluated at the ice-line ($T\simeq 170K$) for every trial $\alpha_{\nu}$ (see \textbf{Fig. 4}).

\section*{Acknowledgements}
\indent We extend our gratitude to Mathieu Roskosz for insightful discussion of the caveats of our study, in particular the potentially significant role water vapor may have played in oxidizing planetesimal interiors. We also thank two anonymous reviewers for constructive reviews that improved the manuscript, and editor Brian Jackson for prompt and careful editorial handling. This work was supported by a Caltech Center for Comparative Planetary Evolution (3CPE) grant to the authors, and a Packard Fellowship to FLHT. 

\section*{References}

Adachi, I., Hayashi, C., \& Nakazawa, K. (1976). The gas drag effect on the elliptic motion of a solid body in the primordial solar nebula. \textit{Progress of Theoretical Physics}, \textit{56}(6), 1756-1771.\\

Arredondo, A., McAdam, M. M., Honniball, C. I., Becker, T. M., Emery, J. P., Rivkin, A. S., ... \& Thomas, C. A. (2024). Detection of Molecular H2O on Nominally Anhydrous Asteroids. \textit{The Planetary Science Journal}, \textit{5}(2), 37.\\

Armitage, P. J. (2020). Astrophysics of planet formation. Cambridge University Press.\\

Batygin, K., \& Morbidelli, A. (2022). Self-consistent model for dust-gas coupling in protoplanetary disks. \textit{Astronomy \& Astrophysics}, \textit{666}, A19.\\

Batygin, K., \& Morbidelli, A. (2023). Formation of rocky super-earths from a narrow ring of planetesimals. \textit{Nature Astronomy}, \textit{7}(3), 330-338.\\

Benz, W., \& Asphaug, E. (1999). Catastrophic disruptions revisited. Icarus, 142(1), 5-20.\\

Bertrand, N., Desgranges, C., Poquillon, D., Lafont, M. C., \& Monceau, D. (2010). Iron oxidation at low temperature (260–500 C) in air and the effect of water vapor. \textit{Oxidation of metals}, \textit{73}(1), 139-162.\\

Birnstiel, T., Klahr, H., \& Ercolano, B. (2012). A simple model for the evolution of the dust population in protoplanetary disks. Astronomy \& Astrophysics, 539, A148.\\

Birnstiel, T. (2023). Dust growth and evolution in protoplanetary disks. \textit{Annual Review of Astronomy and Astrophysics}, \textit{62}.\\

Bitsch, B., Morbidelli, A., Lega, E., \& Crida, A. (2014). Stellar irradiated discs and implications on migration of embedded planets-II. Accreting-discs. \textit{Astronomy \& Astrophysics}, \textit{564}, A135.\\

Bland, P. A., Jackson, M. D., Coker, R. F., Cohen, B. A., Webber, J. B. W., Lee, M. R., ... \& Benedix, G. K. (2009). Why aqueous alteration in asteroids was isochemical: High porosity $\neq$ high permeability. \textit{Earth and Planetary Science Letters}, \textit{287}(3-4), 559-568.\\

Blandford, R. D., \& Payne, D. G. (1982). Hydromagnetic flows from accretion discs and the production of radio jets. \textit{Monthly Notices of the Royal Astronomical Society}, \textit{199}(4), 883-903.\\

Blum, J., \& Wurm, G. (2008). The growth mechanisms of macroscopic bodies in protoplanetary disks. \textit{Annu. Rev. Astron. Astrophys.}, \textit{46}, 21-56.\\

Bonnand, P., \& Halliday, A. N. (2018). Oxidized conditions in iron meteorite parent bodies. \textit{Nature Geoscience}, \textit{11}(6), 401-404.\\

Borlina, C. S., Weiss, B. P., Bryson, J. F., \& Armitage, P. J. (2022). Lifetime of the outer solar system nebula from carbonaceous chondrites. \textit{Journal of Geophysical Research: Planets}, \textit{127}(7), e2021JE007139.\\

Brasser, R., \& Mojzsis, S. J. (2020). The partitioning of the inner and outer Solar System by a structured protoplanetary disk. \textit{Nature Astronomy}, \textit{4}(5), 492-499.\\

Budde, G., Burkhardt, C., Brennecka, G. A., Fischer-Gödde, M., Kruijer, T. S., \& Kleine, T. (2016). Molybdenum isotopic evidence for the origin of chondrules and a distinct genetic heritage of carbonaceous and non-carbonaceous meteorites. \textit{Earth and Planetary Science Letters}, \textit{454}, 293-303.\\

Burkhardt, C., Dauphas, N., Hans, U., Bourdon, B., \& Kleine, T. (2019). Elemental and isotopic variability in solar system materials by mixing and processing of primordial disk reservoirs. \textit{Geochimica et Cosmochimica Acta}, \textit{261}, 145-170.\\

Burkhardt, C., Spitzer, F., Morbidelli, A., Budde, G., Render, J. H., Kruijer, T. S., \& Kleine, T. (2021). Terrestrial planet formation from lost inner solar system material. \textit{Science advances}, \textit{7}(52), eabj7601.\\

Busarev, V. V. (2002). Hydrated silicates on M-, S-, and E-type asteroids as possible traces of collisions with bodies from the Jupiter growth zone. \textit{Solar System Research}, \textit{36}(1), 35-42.\\

Carr, J. S., Tokunaga, A. T., \& Najita, J. (2004). Hot H2O emission and evidence for turbulence in the disk of a young star. \textit{The Astrophysical Journal}, \textit{603}(1), 213.\\

Chabot, N. L. (2004). Sulfur contents of the parental metallic cores of magmatic iron meteorites. \textit{Geochimica et Cosmochimica Acta}, \textit{68}(17), 3607-3618.\\

Chambers, J. E. (2009). An analytic model for the evolution of a viscous, irradiated disk. \textit{The Astrophysical Journal}, \textit{705}(2), 1206.\\

Charnoz, S., Aléon, J., Chaussidon, M., Sossi, P., \& Marrocchi, Y. (2024). Non-Equilibrium Condensation of the Solar Nebula: Forming the Mineralogical Precursors of Chondrites. \textit{LPI Contributions}, \textit{3036}, 6065.\\

Ciesla, F. J., \& Cuzzi, J. N. (2006). The evolution of the water distribution in a viscous protoplanetary disk. \textit{Icarus}, \textit{181}(1), 178-204.\\

Consolmagno, G. J., Britt, D. T., \& Macke, R. J. (2008). The significance of meteorite density and porosity. \textit{Geochemistry}, \textit{68}(1), 1-29.\\

Dauphas, N., \& Chaussidon, M. (2011). A perspective from extinct radionuclides on a young stellar object: the Sun and its accretion disk. \textit{Annual Review of Earth and Planetary Sciences}, \textit{39}, 351-386.\\

Dauphas, N. (2017). The isotopic nature of the Earth’s accreting material through time. \textit{Nature}, \textit{541}(7638), 521-524.\\

Dauphas, N., Hopp, T., \& Nesvorný, D. (2024). Bayesian inference on the isotopic building blocks of Mars and Earth. \textit{Icarus}, \textit{408}, 115805.\\

Delbo’, M., Walsh, K., Bolin, B., Avdellidou, C., \& Morbidelli, A. (2017). Identification of a primordial asteroid family constrains the original planetesimal population. \textit{Science}, \textit{357}(6355), 1026-1029.\\

Donev, A., Cisse, I., Sachs, D., Variano, E. A., Stillinger, F. H., Connelly, R., ... \& Chaikin, P. M. (2004). Improving the density of jammed disordered packings using ellipsoids. \textit{Science}, \textit{303}(5660), 990-993.\\

Dubrulle, B., Morfill, G., \& Sterzik, M. (1995). The dust subdisk in the protoplanetary nebula. \textit{icarus}, \textit{114}(2), 237-246.\\

Dullemond, C. P., Birnstiel, T., Huang, J., Kurtovic, N. T., Andrews, S. M., Guzmán, V. V., ... \& Ricci, L. (2018). The disk substructures at high angular resolution project (DSHARP). VI. Dust trapping in thin-ringed protoplanetary disks. \textit{The Astrophysical Journal Letters}, \textit{869}(2), L46.\\

Dullien, F. A. (2012). \textit{Porous media: fluid transport and pore structure}. Academic press.\\

Ebel, D. S., \& Grossman, L. (2000). Condensation in dust-enriched systems. \textit{Geochimica et Cosmochimica Acta}, \textit{64}(2), 339-366.\\

Feistel, R., \& Wagner, W. (2007). Sublimation pressure and sublimation enthalpy of H2O ice Ih between 0 and 273.16 K. \textit{Geochimica et Cosmochimica Acta}, \textit{71}(1), 36-45.\\

Flaherty, K. M., Hughes, A. M., Rose, S. C., Simon, J. B., Qi, C., Andrews, S. M., ... \& Bai, X. N. (2017). A three-dimensional view of turbulence: constraints on turbulent motions in the HD 163296 protoplanetary disk using DCO+. \textit{The Astrophysical Journal}, \textit{843}(2), 150.\\

Flaherty, K., Hughes, A. M., Simon, J. B., Qi, C., Bai, X. N., Bulatek, A., ... \& Kóspál, Á. (2020). Measuring turbulent motion in planet-forming disks with ALMA: a detection around DM Tau and nondetections around MWC 480 and V4046 Sgr. \textit{The Astrophysical Journal}, \textit{895}(2), 109.\\

Floss, C., Stadermann, F. J., Kearsley, A. T., Burchell, M. J., \& Ong, W. J. (2013). The abundance of presolar grains in comet 81P/Wild 2. \textit{The Astrophysical Journal}, \textit{763}(2), 140.\\

Fry, A., Osgerby, S., \& Wright, M. (2002). Oxidation of alloys in steam environments-a review.\\

Gerbig, K., \& Li, R. (2023). Planetesimal Initial Mass Functions Following Diffusion-regulated Gravitational Collapse. \textit{The Astrophysical Journal}, \textit{949}(2), 81.\\

Goldstein, J. I., Scott, E. R. D., \& Chabot, N. L. (2009). Iron meteorites: Crystallization, thermal history, parent bodies, and origin. \textit{Geochemistry}, \textit{69}(4), 293-325.\\

Greenberg, R., Wacker, J. F., Hartmann, W. K., \& Chapman, C. R. (1978). Planetesimals to planets: Numerical simulation of collisional evolution. \textit{Icarus}, \textit{35}(1), 1-26.\\

Grewal, D. S., \& Asimow, P. D. (2023). Origin of the superchondritic carbon/nitrogen ratio of the bulk silicate Earth–an outlook from iron meteorites. \textit{Geochimica et Cosmochimica Acta}, \textit{344}, 146-159.\\

Grewal, D. S., Nie, N. X., Zhang, B., Izidoro, A., \& Asimow, P. D. (2024). Accretion of the earliest inner Solar System planetesimals beyond the water snowline. \textit{Nature Astronomy}, 1-8.\\

Grossman, L., Fedkin, A. V., \& Simon, S. B. (2012). Formation of the first oxidized iron in the solar system. \textit{Meteoritics \& Planetary Science}, \textit{47}(12), 2160-2169.\\

Gundlach, B., Schmidt, K. P., Kreuzig, C., Bischoff, D., Rezaei, F., Kothe, S., ... \& Stoll, E. (2018). The tensile strength of ice and dust aggregates and its dependence on particle properties. \textit{Monthly Notices of the Royal Astronomical Society}, \textit{479}(1), 1273-1277.\\

Hayashi, C. (1981). Structure of the solar nebula, growth and decay of magnetic fields and effects of magnetic and turbulent viscosities on the nebula. \textit{Progress of Theoretical Physics Supplement}, \textit{70}, 35-53.\\

Hellmann, J. L., Van Orman, J. A., \& Kleine, T. (2024). Hf-W isotope systematics of enstatite chondrites: Parent body chronology and origin of Hf-W fractionations among chondritic meteorites. \textit{Earth and Planetary Science Letters}, \textit{626}, 118518.\\

Hevey, P. J., \& Sanders, I. S. (2006). A model for planetesimal meltdown by 26Al and its implications for meteorite parent bodies. \textit{Meteoritics \& Planetary Science}, \textit{41}(1), 95-106.\\

Hilton, C. D., Ash, R. D., \& Walker, R. J. (2022). Chemical characteristics of iron meteorite parent bodies. \textit{Geochimica et Cosmochimica Acta}, \textit{318}, 112-125.\\

Hunt, A. C., Cook, D. L., Lichtenberg, T., Reger, P. M., Ek, M., Golabek, G. J., \& Schönbächler, M. (2018). Late metal–silicate separation on the IAB parent asteroid: constraints from combined W and Pt isotopes and thermal modelling. \textit{Earth and Planetary Science Letters}, \textit{482}, 490-500.\\

Izidoro, A., Haghighipour, N., Winter, O. C., \& Tsuchida, M. (2014). Terrestrial planet formation in a protoplanetary disk with a local mass depletion: A successful scenario for the formation of Mars. \textit{The Astrophysical Journal}, \textit{782}(1), 31\\

Jacobsen, B., Yin, Q. Z., Moynier, F., Amelin, Y., Krot, A. N., Nagashima, K., ... \& Palme, H. (2008). 26Al–26Mg and 207Pb–206Pb systematics of Allende CAIs: Canonical solar initial 26Al/27Al ratio reinstated. \textit{Earth and Planetary Science Letters}, \textit{272}(1-2), 353-364.\\

Klahr, H., \& Schreiber, A. (2020). Turbulence sets the length scale for planetesimal formation: Local 2D simulations of streaming instability and planetesimal formation. \textit{The Astrophysical Journal}, \textit{901}(1), 54.\\

Kleine, T., Steller, T., Burkhardt, C., \& Nimmo, F. (2023). An inner solar system origin of volatile elements in Mars. \textit{Icarus}, \textit{397}, 115519.\\

Kleine, T., Nimmo, F., Morbidelli, A., \& Nesvorny, D. (2024). Tracing the Provenance of Earth's Accreted Materials from Refractory and Volatile Elements. \textit{LPI Contributions}, \textit{3040}, 2537.\\

Klinkenberge, L. J. (1941). The permeability of porous media to liquids and gases. \textit{Drilling and production practice}, 200-213.\\

Kokubo, E., \& Ida, S. (1998). Oligarchic growth of protoplanets. \textit{Icarus}, \textit{131}(1), 171-178.\\

Kruijer, T. S., Touboul, M., Fischer-Gödde, M., Bermingham, K. R., Walker, R. J., \& Kleine, T. (2014). Protracted core formation and rapid accretion of protoplanets. \textit{Science}, \textit{344}(6188), 1150-1154.\\

Kruijer, T. S., Burkhardt, C., Budde, G., \& Kleine, T. (2017). Age of Jupiter inferred from the distinct genetics and formation times of meteorites. \textit{Proceedings of the National Academy of Sciences}, \textit{114}(26), 6712-6716.\\

Lambrechts, M., \& Johansen, A. (2012). Rapid growth of gas-giant cores by pebble accretion. \textit{Astronomy \& Astrophysics}, \textit{544}, A32.\\

Leinhardt, Z. M., \& Stewart, S. T. (2009). Full numerical simulations of catastrophic small body collisions. \textit{Icarus}, \textit{199}(2), 542-559.\\

Lichtenberg, T., Drażkowska, J., Schönbächler, M., Golabek, G. J., \& Hands, T. O. (2021). Bifurcation of planetary building blocks during Solar System formation. \textit{Science}, \textit{371}(6527), 365-370.\\

Liu, W., Zhang, Y., Tissot, F. L., Avice, G., Ye, Z., \& Yin, Q. Z. (2023). I/Pu reveals Earth mainly accreted from volatile-poor differentiated planetesimals. \textit{Science Advances}, \textit{9}(27), eadg9213.\\

Lynden-Bell, D., \& Pringle, J. E. (1974). The evolution of viscous discs and the origin of the nebular variables. Monthly Notices of the Royal Astronomical Society, 168(3), 603-637.\\

MacPherson, G. J., Davis, A. M., \& Zinner, E. K. (1995). The distribution of aluminum‐26 in the early solar system—A reappraisal. \textit{Meteoritics}, \textit{30}(4), 365-386.\\

Marty, B. (2012). The origins and concentrations of water, carbon, nitrogen and noble gases on Earth. \textit{Earth and Planetary Science Letters}, \textit{313}, 56-66.\\

McCubbin, F. M., \& Barnes, J. J. (2019). Origin and abundances of H2O in the terrestrial planets, Moon, and asteroids. \textit{Earth and Planetary Science Letters}, \textit{526}, 115771.\\

Morbidelli, A., Libourel, G., Palme, H., Jacobson, S. A., \& Rubie, D. C. (2020). Subsolar Al/Si and Mg/Si ratios of non-carbonaceous chondrites reveal planetesimal formation during early condensation in the protoplanetary disk. \textit{Earth and Planetary Science Letters}, \textit{538}, 116220.\\

Morbidelli, A., Kleine, T., \& Nimmo, F. (2025). Did the terrestrial planets of the solar system form by pebble accretion?. \textit{Earth and Planetary Science Letters}, \textit{650}, 119120.\\

Musiolik, G., \& Wurm, G. (2019). Contacts of water ice in protoplanetary disks—laboratory experiments. \textit{The Astrophysical Journal}, \textit{873}(1), 58.\\

Nakagawa, Y., Sekiya, M., \& Hayashi, C. (1986). Settling and growth of dust particles in a laminar phase of a low-mass solar nebula. Icarus, 67(3), 375-390.\\

Nakamoto, T., \& Nakagawa, Y. (1994). Formation, early evolution, and gravitational stability of protoplanetary disks. Astrophysical Journal, Part 1 (ISSN 0004-637X), vol. 421, no. 2, p. 640-650, 421, 640-650.\\

Newcombe, M. E., Nielsen, S. G., Peterson, L. D., Wang, J., Alexander, C. O. D., Sarafian, A. R., ... \& Irving, A. J. (2023). Degassing of early-formed planetesimals restricted water delivery to Earth. \textit{Nature}, \textit{615}(7954), 854-857.\\

Nguyen, A. N., Mane, P., Keller, L. P., Piani, L., Abe, Y., Aléon, J., ... \& Yurimoto, H. (2023). Abundant presolar grains and primordial organics preserved in carbon-rich exogenous clasts in asteroid Ryugu. \textit{Science advances}, \textit{9}(28), eadh1003.\\

Nimmo, F., Kleine, T., Morbidelli, A., \& Nesvorny, D. (2024). Mechanisms and timing of carbonaceous chondrite delivery to the Earth. \textit{Earth and Planetary Science Letters}, \textit{648}, 119112.\\

O’Brien, D. P., Izidoro, A., Jacobson, S. A., Raymond, S. N., \& Rubie, D. C. (2018). The delivery of water during terrestrial planet formation. \textit{Space Science Reviews}, \textit{214}, 1-24.\\

Olson, P., \& Sharp, Z. D. (2018). Hydrogen and helium ingassing during terrestrial planet accretion. \textit{Earth and Planetary Science Letters}, \textit{498}, 418-426.\\

Ormel, C. W., \& Cuzzi, J. N. (2007). Closed-form expressions for particle relative velocities induced by turbulence. Astronomy \& Astrophysics, 466(2), 413-420.\\

Ormel, C. W., \& Klahr, H. H. (2010). The effect of gas drag on the growth of protoplanets-analytical expressions for the accretion of small bodies in laminar disks. \textit{Astronomy \& Astrophysics}, \textit{520}, A43.\\

Ormel, C. W. (2017). The emerging paradigm of pebble accretion. \textit{Formation, Evolution, and Dynamics of Young Solar Systems}, 197-228.\\

Qin, L., Dauphas, N., Wadhwa, M., Masarik, J., \& Janney, P. E. (2008). Rapid accretion and differentiation of iron meteorite parent bodies inferred from 182Hf–182W chronometry and thermal modeling. \textit{Earth and Planetary Science Letters}, \textit{273}(1-2), 94-104.\\

Raymond, S. N., O’Brien, D. P., Morbidelli, A., \& Kaib, N. A. (2009). Building the terrestrial planets: Constrained accretion in the inner Solar System. \textit{Icarus}, \textit{203}(2), 644-662.\\

Ricci, L., Testi, L., Natta, A., Neri, R., Cabrit, S., \& Herczeg, G. J. (2010). Dust properties of protoplanetary disks in the Taurus-Auriga star forming region from millimeter wavelengths. \textit{Astronomy \& Astrophysics}, \textit{512}, A15.\\

Rivkin, A. S., Howell, E. S., Britt, D. T., Lebofsky, L. A., Nolan, M. C., \& Branston, D. D. (1995). 3-$\mu$m spectrophotometric survey of M-and E-class asteroids. \textit{Icarus}, \textit{117}(1), 90-100.\\

Rosotti, G. P., Teague, R., Dullemond, C., Booth, R. A., \& Clarke, C. J. (2020). The efficiency of dust trapping in ringed protoplanetary discs. \textit{Monthly Notices of the Royal Astronomical Society}, \textit{495}(1), 173-181.\\

Rosotti, G. P. (2023). Empirical constraints on turbulence in proto-planetary discs. New Astronomy Reviews, 101674.\\

Rozitis, B., Ryan, A. J., Emery, J. P., Christensen, P. R., Hamilton, V. E., Simon, A. A., ... \& Lauretta, D. S. (2020). Asteroid (101955) Bennu’s weak boulders and thermally anomalous equator. \textit{Science Advances}, \textit{6}(41), eabc3699.\\

Rubin, A. E. (2018). Carbonaceous and noncarbonaceous iron meteorites: Differences in chemical, physical, and collective properties. \textit{Meteoritics \& Planetary Science}, \textit{53}(11), 2357-2371.\\

Safronov, V. S. (1972). Evolution of the Protoplanetary Cloud and Formation of the Earth and the Planets.\\

Scott, G. D., \& Kilgour, D. M. (1969). The density of random close packing of spheres. \textit{Journal of Physics D: Applied Physics}, \textit{2}(6), 863.\\

Sengers, J. V., \& Kamgar-Parsi, B. (1984). \textit{Representative equations for the viscosity of water substance}. American Chemical Society and the American Institute of Physics for the National Bureau of Standards.\\

Sengupta, D., Cuzzi, J. N., Umurhan, O. M., \& Lyra, W. (2024). Length and Velocity Scales in Protoplanetary Disk Turbulence. \textit{The Astrophysical Journal}, \textit{966}(1), 90.\\

Shakura, N. I., \& Sunyaev, R. A. (1973). Black holes in binary systems. Observational appearance. Astronomy and Astrophysics, Vol. 24, p. 337-355, 24, 337-355.\\

Sossi, P. A., Stotz, I. L., Jacobson, S. A., Morbidelli, A., \& O’Neill, H. S. C. (2022). Stochastic accretion of the Earth. \textit{Nature astronomy}, \textit{6}(8), 951-960.\\

Spitzer, F., Burkhardt, C., Nimmo, F., \& Kleine, T. (2021). Nucleosynthetic Pt isotope anomalies and the Hf-W chronology of core formation in inner and outer solar system planetesimals. \textit{Earth and Planetary Science Letters}, \textit{576}, 117211.\\

Squire, J., \& Hopkins, P. F. (2018). Resonant drag instabilities in protoplanetary discs: the streaming instability and new, faster growing instabilities. \textit{Monthly Notices of the Royal Astronomical Society}, \textit{477}(4), 5011-5040.\\

Tabone, B., Rosotti, G. P., Cridland, A. J., Armitage, P. J., \& Lodato, G. (2022). Secular evolution of MHD wind-driven discs: analytical solutions in the expanded $\alpha$-framework. Monthly Notices of the Royal Astronomical Society, 512(2), 2290-2309.\\

Testi, L., Natta, A., Shepherd, D. S., \& Wilner, D. J. (2003). Large grains in the disk of CQ Tau. \textit{Astronomy \& Astrophysics}, \textit{403}(1), 323-328.\\

Tissot, F. L., Collinet, M., Namur, O., \& Grove, T. L. (2022). The case for the angrite parent body as the archetypal first-generation planetesimal: Large, reduced and Mg-enriched. \textit{Geochimica et Cosmochimica Acta}, \textit{338}, 278-301.\\

Trapman, L., Rosotti, G., Bosman, A. D., Hogerheijde, M. R., \& Van Dishoeck, E. F. (2020). Observed sizes of planet-forming disks trace viscous spreading. \textit{Astronomy \& Astrophysics}, \textit{640}, A5.\\

Urey, H. C., \& Craig, H. (1953). The composition of the stone meteorites and the origin of the meteorites. \textit{Geochimica et Cosmochimica Acta}, \textit{4}(1-2), 36-82.\\

Villenave, M., Ménard, F., Dent, W. R. F., Duchêne, G., Stapelfeldt, K. R., Benisty, M., ... \& Padgett, D. (2020). Observations of edge-on protoplanetary disks with ALMA-I. Results from continuum data. \textit{Astronomy \& Astrophysics}, \textit{642}, A164.\\

Vorobyov, E. I. (2009). Variable accretion in the embedded phase of star formation. \textit{The Astrophysical Journal}, \textit{704}(1), 715.\\

Wang, H., Weiss, B. P., Bai, X. N., Downey, B. G., Wang, J., Wang, J., ... \& Zucolotto, M. E. (2017). Lifetime of the solar nebula constrained by meteorite paleomagnetism. \textit{Science}, \textit{355}(6325), 623-627.\\

Warren, P. H. (2011). Stable-isotopic anomalies and the accretionary assemblage of the Earth and Mars: A subordinate role for carbonaceous chondrites. \textit{Earth and Planetary Science Letters}, \textit{311}(1-2), 93-100.\\

Weidenschilling, S. J., \& Cuzzi, J. N. (1993). Formation of planetesimals in the solar nebula. In \textit{Protostars and planets III} (pp. 1031-1060).\\

Weiss, B. P., Bai, X. N., \& Fu, R. R. (2021). History of the solar nebula from meteorite paleomagnetism. \textit{Science Advances}, \textit{7}(1), eaba5967.\\

Yap, T. E., \& Batygin, K. (2024). Dust-gas coupling in turbulence-and MHD wind-driven protoplanetary disks: Implications for rocky planet formation. \textit{Icarus}, \textit{417}, 116085.\\

Yap, T. E., \& Tissot, F. L. (2023). The NC-CC dichotomy explained by significant addition of CAI-like dust to the Bulk Molecular Cloud (BMC) composition. \textit{Icarus}, \textit{405}, 115680.\\

Youdin, A. N., \& Goodman, J. (2005). Streaming instabilities in protoplanetary disks. The Astrophysical Journal, 620(1), 459.\\

Yuan, J., Wang, W., Zhu, S., \& Wang, F. (2013). Comparison between the oxidation of iron in oxygen and in steam at 650–750 C. \textit{Corrosion Science}, \textit{75}, 309-317.\\

Zhang, W., Thompson, K. E., Reed, A. H., \& Beenken, L. (2006). Relationship between packing structure and porosity in fixed beds of equilateral cylindrical particles. \textit{Chemical Engineering Science}, \textit{61}(24), 8060-8074.\\

Zhang, Z. (2023). Ice Sublimation in Planetesimals Formed at the Outward Migrating Snowline. \textit{The Astrophysical Journal Letters}, \textit{956}(1), L25.\\
\end{document}